\documentclass[fleqn,usenatbib]{mnras}

\usepackage{newtxtext,newtxmath}

\usepackage[T1]{fontenc}

\usepackage{graphicx}	
\usepackage{xspace}

\newcommand{\integral}{\textit{INTEGRAL}}
\newcommand{\rxte}{\textit{RXTE}}

\newcommand{\nustar}{\textit{NuSTAR}}
\newcommand{\xmm}{\textit{XMM-Newton}}

\newcommand{\kev}{keV}
\newcommand{\frone}{$\Phi_{\mathrm{8-24\ keV}}$\xspace}
\newcommand{\frtwo}{$\Phi_{\mathrm{30-50\ keV}}$\xspace}
\graphicspath{{newIndividualPanels/} {plots/}}
\newcommand{\dPhi}{$\Delta\Phi/\Phi$\xspace}

\title[Coronal Cooling by X-ray Bursts]{Cooling of Accretion Disc Coronae by Type I X-ray Bursts}

\author[J. Speicher et al.]{
J. Speicher,$^{1}$\thanks{E-mail: juliasp@gatech.edu}
D. R. Ballantyne$^{1}$
 and J. Malzac$^{2}$
\\
$^{1}$Center for Relativistic Astrophysics, School of Physics, Georgia
  Institute of Technology, 837 State Street, Atlanta, GA 30332-0430, USA\\
$^{2}$ IRAP, Universit\'{e} de Toulouse, CNRS, UPS, CNES, Toulouse, France.
}

\date{Accepted XXX. Received YYY; in original form ZZZ}

\pubyear{2020}

\begin{document}
\label{firstpage}
\pagerange{\pageref{firstpage}--\pageref{lastpage}}
\maketitle

\begin{abstract}
Although accretion disc coronae appear to be common in many accreting
systems, their fundamental properties remain insufficiently
understood. Recent work suggests that Type I X-ray bursts from
accreting neutron stars provide an opportunity to probe the
characteristics of coronae. Several studies have observed hard X-ray
shortages from the accretion disk during an X-ray burst implying
strong coronal cooling by burst photons. Here, we use the plasma
emission code \textsc{eqpair} to study the impact of X-ray bursts on
coronae, and how the coronal and burst properties affect the coronal
electron temperatures and emitted spectra. Assuming a constant
accretion rate during the burst, our simulations show that soft
photons can cool coronal electrons by a factor of $\gtrsim 10$ and cause a
reduction of emission in the $30$--$50$~keV band to $\lesssim 1\%$ of the
pre-burst emission. This hard X-ray drop is intensified when the
coronal optical depth and aspect ratio is increased. In contrast,
depending on the properties of the burst and corona, the emission in
the $8$--$24$ keV band can either increase, by a factor of $\gtrsim20$, or
decrease, down to $\lesssim 1\%$ of the pre-burst emission. An
increasing accretion rate during the X-ray burst reduces the coronal
cooling effects and the electron temperature drop can be mitigated by
$\gtrsim60\%$. These results indicate that changes of the hard X-ray flux during an X-ray burst probe the geometrical properties of the corona.
\end{abstract}

\begin{keywords}
accretion, accretion discs -- radiation mechanisms: thermal -- stars:
neutron -- X-rays: binaries -- X-rays: bursts
\end{keywords}



\section{Introduction}
\label{sect:intro}

Hard X-ray emission is regularly detected from accreting compact
objects (i.e., neutron stars and black holes), and is thought to
originate in a corona, a region of hot electrons located close to the accretion flow where soft photons from the disc are Compton
up-scattered to X-ray energies
\citep[e.g.,][]{Thorne1975,Galeev1979,White1982,Zdziarski2004,Ibragimov2005,Revnivtsev2008,Wang2019}.
Despite their near ubiquity, the
connection between accretion disc properties (e.g., accretion rate)
and the characteristics of the corona (e.g., size and optical
depth) is not well understood. For example, several models and
observations indicate that when the accretion rate is low the corona
takes up the region between the central object and a truncated thin disc
(e.g., \citealt{Gierlinski2002,Mayer2007}; for a review see
\citealt{Done2007}). However, direct observational tests of the geometry
of the corona remain elusive.

While both neutron stars and black holes are accreting compact
objects, the observational properties of their corona appear to be
significantly different. For example, for a given optical depth, the
measured electron temperatures of coronae in accreting neutron stars are
lower than what is measured for black holes systems, leading to
steeper X-ray spectra \citep{Burke2017}. Radiation emitted
from the neutron star surface and the disc boundary layer provide
additional sources of soft photons that can interact and cool the corona.

An accreting neutron star (NS) can also release a large number of soft photons
through a Type I X-ray burst, which occurs after the star has accreted
sufficient hydrogen and helium from a
companion star that has filled its Roche lobe
\citep[e.g.,][]{Lewin1993, Cumming2003}. The accreted matter
accumulates onto the neutron star, where the pressure and temperature
rise, leading eventually to unstable nuclear burning of the accreted
layer. The burning front quickly envelopes the star, heating the
stellar surface which is then detectable as an X-ray
burst (generally well described by a blackbody; e.g.,
\citealt{Swank1977,DegenaarEtAl2016}) that lasts, for many bursts, at most a few 10s of
seconds (for a recent review see \citealt{GallowayReview2017}).
The intense, transient burst of soft X-ray photons will then interact
with the surrounding accretion disk and corona, leading to a host of
potentially interesting effects \citep[e.g.,][]{bs04,be05,DegenaarEtAl2018}.


The impact of the burst on the corona is particularly important as it
could change the `persistent spectrum' emitted by the disc and
corona during the burst. Historically, analysis of X-ray burst spectra
commonly assumed the persistent spectrum to
be unchanged from its pre-burst properties, and therefore could be safely
subtracted from the burst spectrum in order to measure burst
properties
\citep[e.g.,][]{Sztajno1986,Galloway2008,Ji2013}. Critically, these
techniques have been used to infer the neutron star radius, which would
constrain the equation-of-state of nuclear matter \citep[e.g.,][]{Lewin1993,Ozel2009,Suleimanov2011}. However, the
assumption that the persistent spectrum is constant throughout the
burst has come into question. \citet{Worpel2013} introduced a
normalization factor for the persistent spectrum and found that it
exceeded unity in a statistical analysis of many \rxte\ burst spectra, suggesting a persistent emission
enhancement \citep[also shown by][]{Worpel2015}. The enhancement could
be caused by an increase in the accretion rate during the burst due to
the loss of angular
momentum from Poynting-Robertson drag
\citep{Walker1989,Walker1992}. This effect has been confirmed by
recent numerical simulations investigating the impact of a burst on a
thin accretion disk \citep{fbb19}.

In addition to the change in the normalization of the persistent
emission, recent observations have found evidence for a hard X-ray
shortage during X-ray bursts. The earliest example was discovered by
\citet{Maccarone2003} who found that the 30-60 keV emission dropped by
a factor of 2 during a burst from Aql~X-1. Stacking of bursts have led
to several measurements with large significance, most recently by
\citet{Sanchez2020} who detected an 80$\%$ emission shortage in the 35-70
keV band after stacking several bursts from GS~1826–238. As these
changes occur at much higher energies than found in the burst
spectrum, they appear to be evidence of changes in the corona due to
the burst, likely from Compton cooling due to the large influx of soft burst photons \citep[e.g.,][]{Maccarone2003,jiEtAl2014KS1731-260,ji2014,DegenaarEtAl2018,Chen2018ApJ,SimulatingtheCollapseofaThickAccretionDisk,Sanchez2020}.
It appears that a close investigation of the hard X-ray flux properties during a burst
may therefore provide an unique probe of the corona.

Hence, this paper explores how an X-ray burst impacts the emission
from an accretion disc corona. We examine the role of both coronal and
burst properties in the coronal cooling process, and also investigate the
effects of an enhanced accretion rate during the burst. The
calculations are described in Section \ref{sect:calc} with the results
presented in Section \ref{sect:results}. Section \ref{sect:discuss}
contains a comparison of our results to observations and discusses
other effects not explicitly considered in the current
calculations. Finally, Section \ref{sect:concl} summarizes the
principle conclusions and key results of this work.

\section{Calculations}
\label{sect:calc}


To assess the impact of the X-ray burst photons on the corona, we will
consider an idealized neutron star system emitting an isotropic X-ray
burst. The neutron star is surrounded by a azimuthally symmetric
corona which connects onto a truncated thin accretion disc. This
scenario was initially presented and described by \citet{DegenaarEtAl2018} (see
their Section 5.2.3), and we build upon their treatment as described below. Since we are
focused on the impact of the burst on the corona, we will ignore the
influence of reprocessing of the burst in the thin disk (the impact of
reflection is considered in Sect.~\ref{sub:reflection}).


Following \citet{DegenaarEtAl2018}, the total luminosity available to
the system is defined as
\begin{equation}\label{eq:totalAccretionL}
L = \frac{G M_* \beta \Dot{M}}{R},
\end{equation}
with $M_*$ being the mass of the neutron star, $\Dot{M}$ the mass
accretion rate and $R$ the neutron star radius. The factor $\beta$ introduced in Eq.~\ref{eq:totalAccretionL} is our first expansion of the calculations by
\citet{DegenaarEtAl2018}. $\beta$ quantifies any enhanced mass
accretion rate that may be initiated by the X-ray burst interacting
with its surroundings \citep[e.g.,][]{Worpel2015,SimulatingtheCollapseofaThickAccretionDisk}. 
For simplicity, we assume $\beta$ is a linear function of the burst
luminosity, $L_{\mathrm{b}}$,
\begin{equation}\label{eq:beta}
\beta = 1 + \frac{1}{5}\frac{L_\mathrm{b}}{L_0},
\end{equation}
where $L_0$ is the unmodified luminosity (i.e., $L_0=G M_* \Dot{M}/R$).
The chosen slope of $1/5$ ensures that the enhancement of the
accretion rate is at most a factor of $5$ for the maximum considered burst luminosity ($L_{\mathrm{b}}/L_0=20$), consistent with the changes
in the normalization of the persistent spectrum
measured in recent X-ray bursts ($\lesssim 5$; \citealt{DegenaarEtAl2016, KeekEtAl2018}).
Section~\ref{sect:results} shows the effect of this increase in mass accretion rate on
the changes in the coronal emission by comparing the results to models
computed with $\beta=1$.

The accretion energy $L$ is distributed through the neutron star
system, with half of the energy released as the luminosity of the disc
$L_{\mathrm{d}}$ and corona $L_{\mathrm{c}}$, and the remainder
emitted by the neutron star $L_{\mathrm{NS}}$, which includes the
emission of accreted matter arriving at the neutron star surface as
well as at its boundary layer \citep[e.g.,][]{Burke2017}. 
If $f$ is the fraction of the accretion energy transported to the corona,
the disc luminosity is defined as
\begin{equation}\label{eq:diskL}
L_\mathrm{d} = (1-f)\frac{L}{2}.
\end{equation}
In addition, a fraction $f_\mathrm{c}$ of the energy is advected through the corona
to the neutron star instead of being dissipated within the
corona. Therefore, the energy dissipation rate in the corona is
\begin{equation}\label{eq:coronaL}
L_\mathrm{c} = (1-f_\mathrm{c})f\frac{L}{2}.
\end{equation}
%
Lastly, the accretion energy that survives to the neutron star is
finally released as
\begin{equation}\label{eq:nsL}
L_{\mathrm{NS}} = L - L_{\mathrm{d}}-L_\mathrm{c} =(1+f_{\mathrm{c}}f)\frac{L}{2}.
\end{equation}
Note that all three of these luminosities are also proportional to $\beta$.

The energy released in the corona, $L_{\mathrm{c}}$, is used to heat
electrons which then Compton up-scatters the soft photons emitted by the
disc and NS to produce the cutoff power-law spectrum observed as the
persistent emission \citep[e.g.,][]{ChenEtAl2012, Kajava2017}. The efficiency of the interaction between the
corona and the seed photons emitted by the disc and NS will depend on
the size and geometry of the corona, as only a certain fraction
($f_\mathrm{d}$ and $f_{\mathrm{NS}}$, respectively) of the
disc and NS emission will enter the corona. Therefore, we define the
soft luminosity entering the corona prior to any X-ray burst as
%
\begin{equation}\label{eq:ls0}
L_\mathrm{s,0} = f_\mathrm{d} L_\mathrm{d} + f_{\mathrm{NS}} L_{\mathrm{NS}}.
\end{equation}
An X-ray burst with luminosity $L_\mathrm{b}$ also radiates from the
NS surface, so during a burst the soft photon luminosity that enters
the corona increases to 
\begin{equation}\label{eq:ls}
L_\mathrm{s} = L_\mathrm{s,0} + f_{\mathrm{NS}} L_\mathrm{b},
\end{equation}
where, following \citet{DegenaarEtAl2018}, we assume the burst and the NS emission are isotropic
point sources as viewed from the corona.
From above, we see that the impact of the soft photons on the corona
depends on $f_\mathrm{c}$, $f$, $f_\mathrm{d}$ and $f_\mathrm{NS}$. We
find that, in
the truncated disc scenario investigated here, only $f_{\mathrm{c}}$ and $f_{\mathrm{NS}}$
have a significant impact on the spectrum produced by the corona during
an X-ray burst. However, the impact of changes in $f_{\mathrm{c}}$ are similar to the ones caused by $\beta \neq 1$. Therefore, following \citet{DegenaarEtAl2018}, we set
$f_\mathrm{c}=0.2$, $f=0.9$, and $f_\mathrm{d}=0.05$. The
$f_{\mathrm{NS}}$ parameter is related to the opening angle of the
corona with respect to the NS and thus depends on the aspect ratio
$\epsilon$, defined by the coronal height $h$ and radius $r$ so that
$\epsilon = h/r$:
\begin{equation}\label{eq:fns}
f_\mathrm{NS} = \frac{\epsilon}{\sqrt{1+\epsilon^2}}.
\end{equation}
A larger $\epsilon$ means that a larger fraction of the burst photons
will interact with the corona. Finally, the optical depth $\tau$ of the
corona will also be important in determining the strength of the
interaction between the burst and the corona, as a larger $\tau$ will
lead to more frequent scatterings for a given $\epsilon$. We therefore
consider both $\epsilon$ and $\tau$ as control variables for our
simulations.

We assume the burst, disc and NS all radiate as blackbodies.
In contrast to \citet{DegenaarEtAl2018}, we treat the blackbody
temperature of the X-ray burst ($T_{\mathrm{b}}$) separately from the temperatures of the disc and NS (which are assumed to be equal).
%
%
The initial temperature of the burst,
$T_{\mathrm{b,0}}$, is a free parameter of the model and corresponds to $L_{\mathrm{NS}}$ with
$\beta=1$ (Eq.~\ref{eq:nsL}). With this definition, the blackbody
luminosity rises to $L_{\mathrm{NS}} + L_\mathrm{b}$ during the burst,
and the blackbody temperature of the burst increases as
\begin{equation}\label{eq:TBB_b}
T_{\mathrm{b}} =  T_{\mathrm{b,0}}\left(\frac{1+f_\mathrm{c} f + 2L_\mathrm{b}/L_0}{1+f_\mathrm{c} f}\right)^{1/4}.
\end{equation}

We utilize the hybrid plasma
emission code \textsc{eqpair} \citep[][]{Coppi1999} to calculate the
equilibrium electron temperature of the corona and its emitted
spectrum for a wide range of burst and coronal properties. The input parameters for
\textsc{eqpair}, $L_\mathrm{c}/L_\mathrm{s}$ and $L_\mathrm{s}/ L$, are
calculated using the equations described above. The seed photon
temperature is set to $kT_{\mathrm{b}}$ (Eq.~\ref{eq:TBB_b}). With this setup we run a grid of
models with $L_{\mathrm{b}}/L_0=10^{-4}\ldots 20$ (in 100 logarithmically
spaced steps) and $kT_{\mathrm{b,0}}=0.1\ldots 0.9$~keV (in 33 linearly
spaced steps). This temperature range yields reasonable maximum
$kT_{\mathrm{b}}$, but we note that higher temperature discs are
observed in some hard-state NS X-ray binaries \citep{Burke2017}.  A model with $L_{\mathrm{b}}/L_0=0$ is also computed as part of
the grid. Every grid is calculated for three different coronal
aspect ratios ($\epsilon=0.2, 0.5$ and $1$), each with two different optical
depths ($\tau=0.5$ and $1.5$). All calculations were run once
where the accretion rate was unchanged during the X-ray burst (i.e.,
$\beta=1$), and once when the accretion power increased during the
burst (Eq.~\ref{eq:beta}). Finally, in accordance with our choice to
focus on the burst-corona interaction, the reflection strength is set
to zero in \textsc{eqpair}, and we restrict our analysis to an
entirely thermal corona\footnote{The addition of a small non-thermal
electron population in the corona has only a small quantitative effect
on the results present below. However, the non-thermal population does
allow the production of a faint annihilation line in the spectrum
\citep{Maciolek-Niedzwiecki1995}. We defer an investigation on the
relationship of the burst to the emitted line to future work.}.



\section{Results}
\label{sect:results}
\subsection{Fixed Accretion Luminosity}
\label{sub:fixedbeta}

\begin{figure*}
    \centering
    \includegraphics[width = 0.47\textwidth]{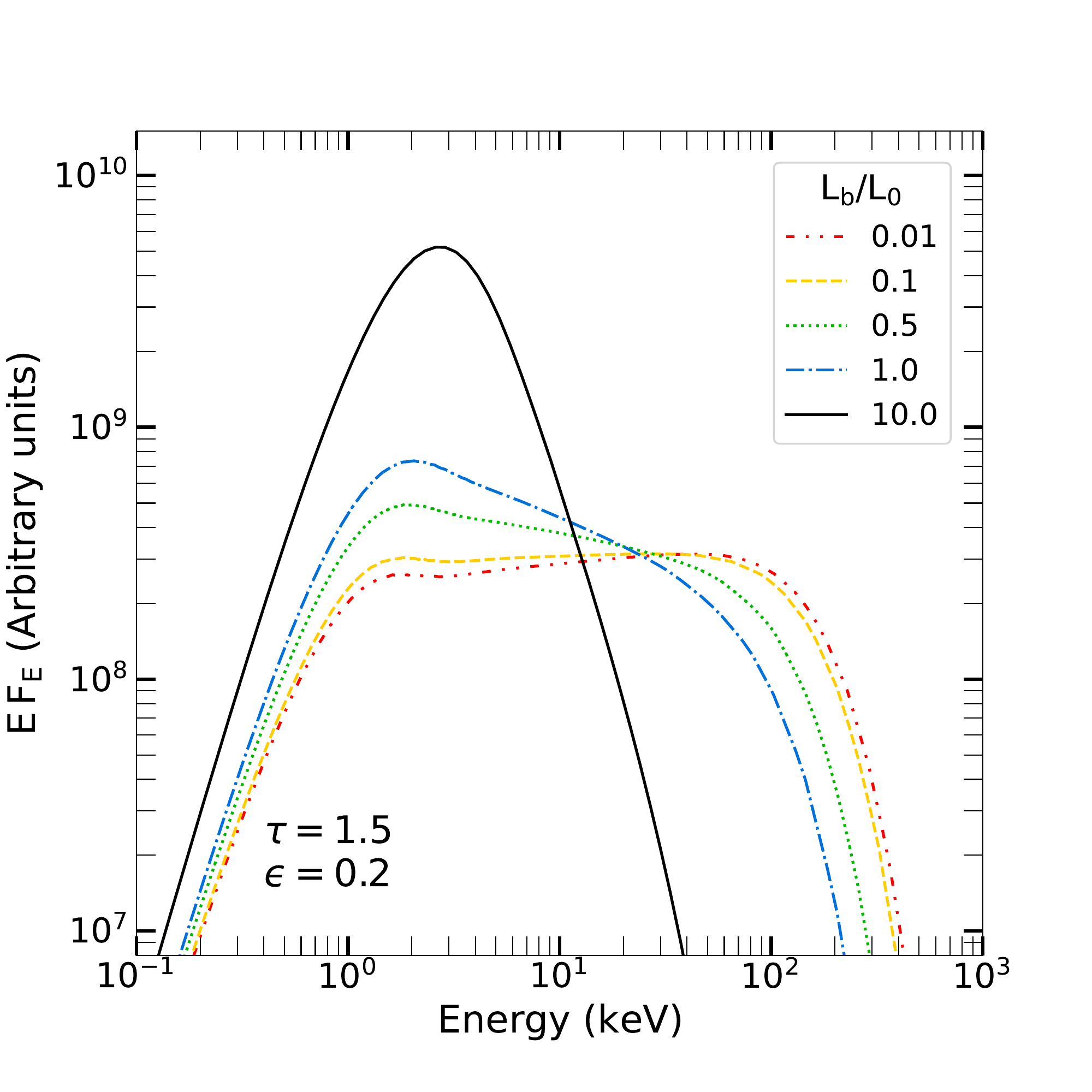}
    \includegraphics[width = 0.47\textwidth]{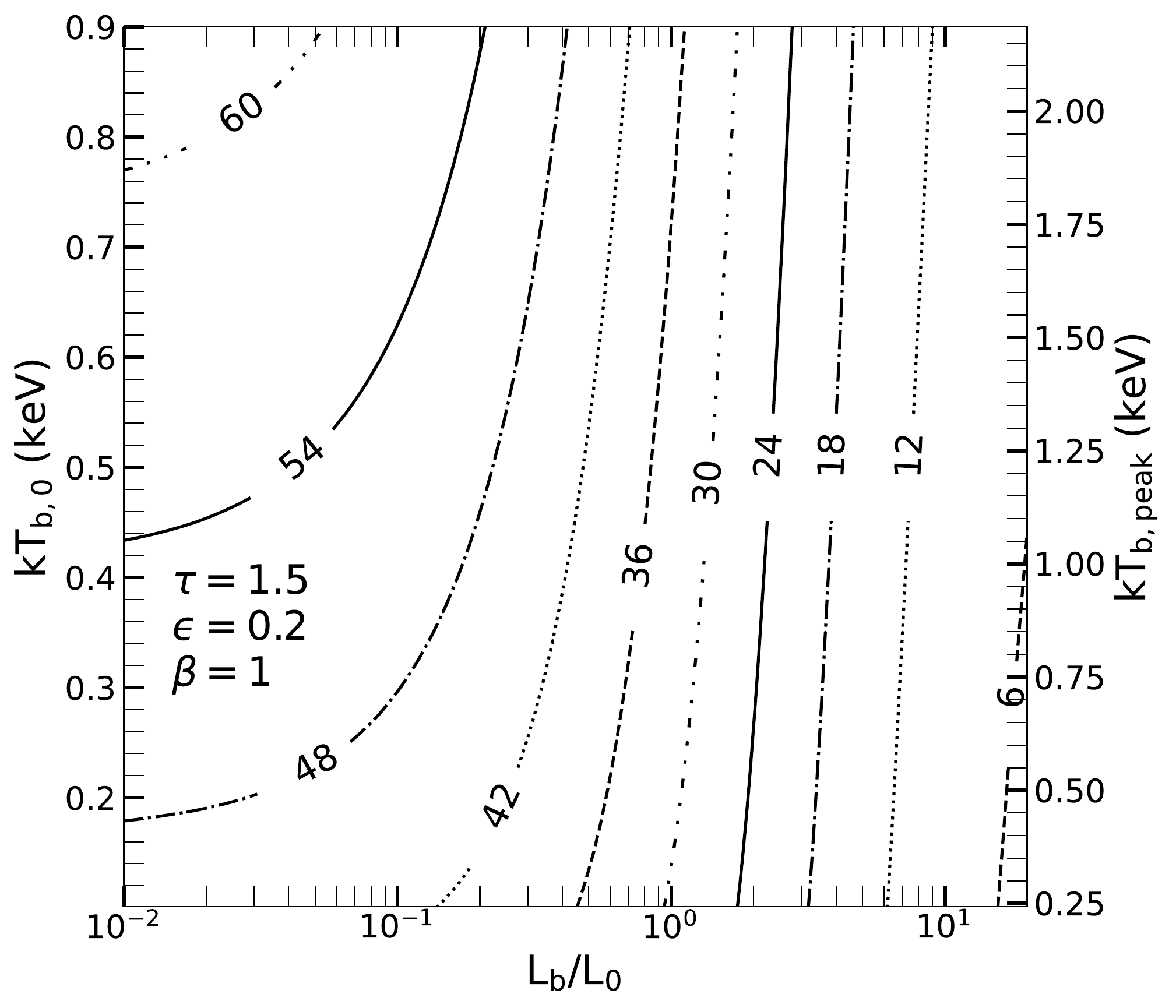}
\caption{(Left) Examples of the X-ray spectra emitted by an accretion
  disc corona irradiated by an X-ray burst with luminosity
  $L_{\mathrm{b}}/L_0$. In these calculations, the corona has an optical
  depth $\tau=1.5$ and an aspect ratio of $\epsilon=0.2$. The initial
  temperature of the burst and NS is $kT_{\mathrm{b,0}}=0.3$~\kev, and
  the accretion luminosity, $L$, does not change during the burst
  (i.e., $\beta=1$). The five lines show the spectra for
  $L_{\mathrm{b}}/L_0=0.01$ (red, double-dotted dashed line), $0.1$
  (yellow dashed line), $0.5$ (green dotted line), $1$ (blue
  dot-dashed line), and $10$ (black solid line). Compton scattering of
  the burst photons in the corona softens the spectra while reducing
  its cutoff energy, with the largest effects at
  $L_{\mathrm{b}}/L_0 \ga 0.1$. (Right) Contours of the coronal electron
  temperature (in units of keV) as a function of the initial burst
  temperature ($kT_{\mathrm{b,0}}$) and burst luminosity
  ($L_{\mathrm{b}}/L_0$). The right-hand axis shows the peak blackbody temperature
  reached by the X-ray burst, $kT_{\mathrm{b,peak}}$. The coronal parameters are the same as in the
  other panel, as is $\beta$. Comptonization of the burst photons
  cools the coronal plasma with the most rapid cooling occurring at
  the largest $kT_{\mathrm{b,0}}$.}
    \label{spectraVaryingLbConstantB}
\end{figure*}

We begin by considering the impact of an X-ray burst on the corona for
the case of a constant accretion luminosity (i.e., $\beta=1$). The left
panel of Figure~\ref{spectraVaryingLbConstantB} shows the effect of an
X-ray burst with luminosity $L_\mathrm{b}/L_0$ = 0.01 (red, double-dotted
dashed line), 0.1 (yellow dashed line), 0.5 (green dotted line),
1 (blue dot-dashed line), and 10 (black solid line) on the spectra
emitted by the corona. For these examples, the initial burst
temperature is $kT_\mathrm{b,0}=0.3$ keV, and the corona has an
optical depth $\tau=1.5$ and aspect ratio $\epsilon=0.2$. A larger
$L_{\mathrm{b}}$ increases the soft photon luminosity
(Eq.~\ref{eq:ls}), but the energy dissipated in the corona remains the
same in each calculation (Eq.~\ref{eq:coronaL}). The increasing
$L_\mathrm{s}$ means more soft burst photons illuminate the corona,
cooling it down via inverse Comptonization which softens the spectrum
and decreases the high-energy cutoff. These changes in the spectrum
are most noticeable when $L_{\mathrm{b}}/L_0 \ga 0.1$ (see also
\citealt{DegenaarEtAl2018}). At $L_{\mathrm{b}}/L_0 \ga 10$ the burst
dominates the spectrum and it becomes overwhelmingly thermal.

The equilibrium temperature of the corona depends on both the initial
burst temperature as well as its luminosity. The right panel of
Fig.~\ref{spectraVaryingLbConstantB} shows contours of the equilibrium
coronal electron temperature when $L_\mathrm{b}/L_0$ is varied from
$10^{-2}$ to 20 and $kT_\mathrm{b,0}$ from 0.1 keV to 0.9 keV. As seen
in the right-hand axis of the panel, this
range of $kT_\mathrm{b,0}$ yields a maximum $kT_\mathrm{b}$ of
$0.24$~keV for $kT_\mathrm{b,0}=0.1$~keV and $2.2$~keV for
$kT_\mathrm{b,0}=0.9$ keV (Eq.~\ref{eq:TBB_b}). The optical depth and
aspect ratio of the corona are the same as in the left panel. The plot
shows that when the burst luminosity is small (i.e., $L_{\mathrm{b}}/L_0
\lesssim 0.1$) the X-ray burst has a minor effect on the corona with
  its temperature largely determined by the seed photon temperature,
  $kT_{\mathrm{b,0}}$, as higher energy photons are less efficient in cooling the corona. However, when the X-ray burst begins to
  become a large fraction of the total soft luminosity (at
  $L_\mathrm{b}/L_0 \ga 0.5$), the coronal temperature becomes largely
  independent from $kT_\mathrm{b,0}$. Indeed, the contours of coronal temperature
  become nearly vertical in the figure and cooling by a factor of $\gtrsim 10$ is possible. Hence, X-ray bursts from hot
  NS surfaces cause the greatest change in the coronal electron
  temperature during the burst.

%
\begin{figure*}
    \centering
    \includegraphics[width = 0.9\textwidth]{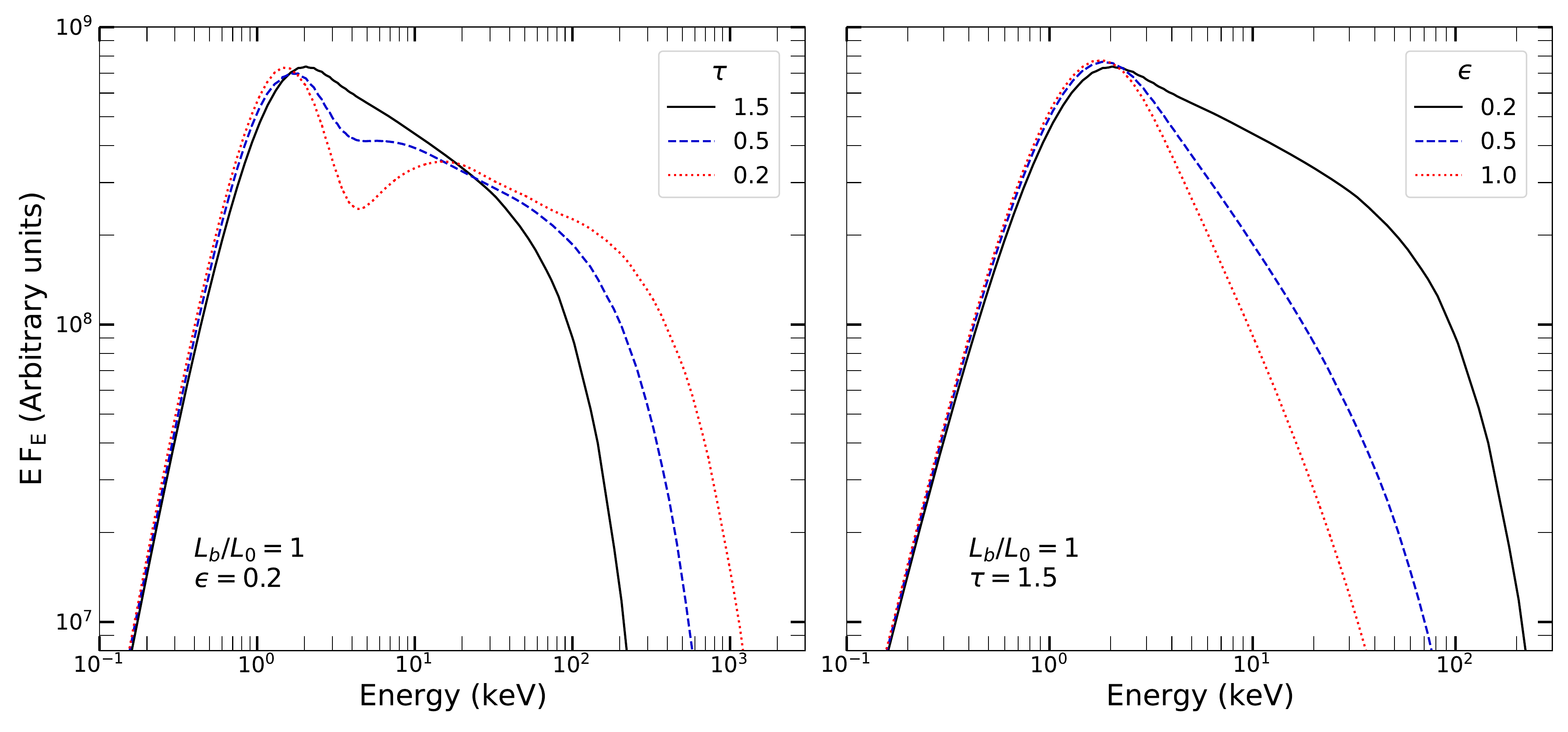}
\caption{(Left) The effect of the coronal optical depth on the emitted
  burst-cooled spectrum. The three lines all denote spectra with
  $L_{\mathrm{b}}/L_0=1$ and aspect ratio $\epsilon=0.2$, and indicate
  optical depths $\tau=1.5$ (black solid line), $0.5$ (blue dashed
  line) and $0.2$ (dotted red line). The initial burst temperature is
  $kT_{\mathrm{b,0}}=0.3$~\kev\ and the accretion luminosity is
  constant throughout the burst (i.e., $\beta=1$;
  Eq.~\ref{eq:beta}). A smaller coronal optical depth reduces the
  cooling effects of the burst in the corona, and the spectra retains
  characteristics of high temperatures. (Right) As in the other panel,
  but now showing the effects of changing the aspect ratio of the
  corona. The optical depth is $\tau=1.5$ in the 3 cases shown here:
  $\epsilon=0.2$ (solid black line), $0.5$ (dashed blue line) and
  $1.0$ (dotted red line). A larger aspect ratio means that a higher
  fraction of the burst photons encounter the corona (Eqs.~\ref{eq:ls}
  and~\ref{eq:fns}) leading to more substantial cooling for the same
  $L_{\mathrm{b}}/L_0$. The hard X-ray fluxes measured during an X-ray burst with a
  given luminosity will depend on both $\tau$ and $\epsilon$.}
    \label{varyTauEpsilonConstantB}
\end{figure*}
The observed hard X-ray flux will depend not only on the burst
luminosity, but also on the coronal optical depth $\tau$ and the
aspect ratio $\epsilon$. The left panel of Figure
\ref{varyTauEpsilonConstantB} illustrates the effect of the coronal
optical depth on the emitted burst-cooled spectrum when
$L_\mathrm{b}/L_0=1$. In these examples, the initial X-ray burst
temperature is set to $kT_\mathrm{b,0}=0.3$ keV and the aspect ratio
of the corona is $\epsilon=0.2$, while $\tau=1.5$ (black solid line),
$0.5$ (blue dashed line) and $0.2$ (dotted red line). A larger $\tau$
increases the likelihood of interactions between the coronal electrons
and soft photons which cool the corona and soften the emitted spectrum. 
The $\tau=0.2$ model is shown here to illustrate the case of an
extremely optically-thin corona, although such a scenario may be
uncommon in real NS systems.
The blackbody shape of the X-ray burst is visible in the $\tau=0.2$
spectrum, but is largely invisible when $\tau=1.5$ and scatterings are
more frequent. Crucially, the softer spectrum produced by the cooler corona
with $\tau=1.5$ strongly impacts the observed hard X-ray flux. A
similar pattern is observed in the right panel of
Fig.~\ref{varyTauEpsilonConstantB}, which illustrates the impact of the
coronal aspect ratio, $\epsilon$, on the emitted spectrum. As
$\epsilon$ sets the amount of burst photons intercepting the corona
(Eq.~\ref{eq:fns}), this parameter will also affect the temperature of
the corona during a burst. The examples shown in the plot are
calculated with $\tau=1.5$ and $\epsilon=0.2$ (solid black line),
$0.5$ (dashed blue line) and $1.0$ (dotted red line). If the corona
provides a large target as seen from the NS, i.e. a large $\epsilon$,
the corona can be efficiently cooled, producing a large hard X-ray
shortage in the spectrum. By comparing these two panels of
Fig.~\ref{varyTauEpsilonConstantB}, it is apparent that the spectrum
emitted by the corona will depend more sensitively on $\epsilon$ than
$\tau$. Therefore, measurements of the hard X-ray flux deficit during
an X-ray burst could constrain the aspect ratio of the accretion disc
corona.


%
%
%
%
%
\begin{figure*}
    \centering
    \includegraphics[width = 0.45\textwidth]{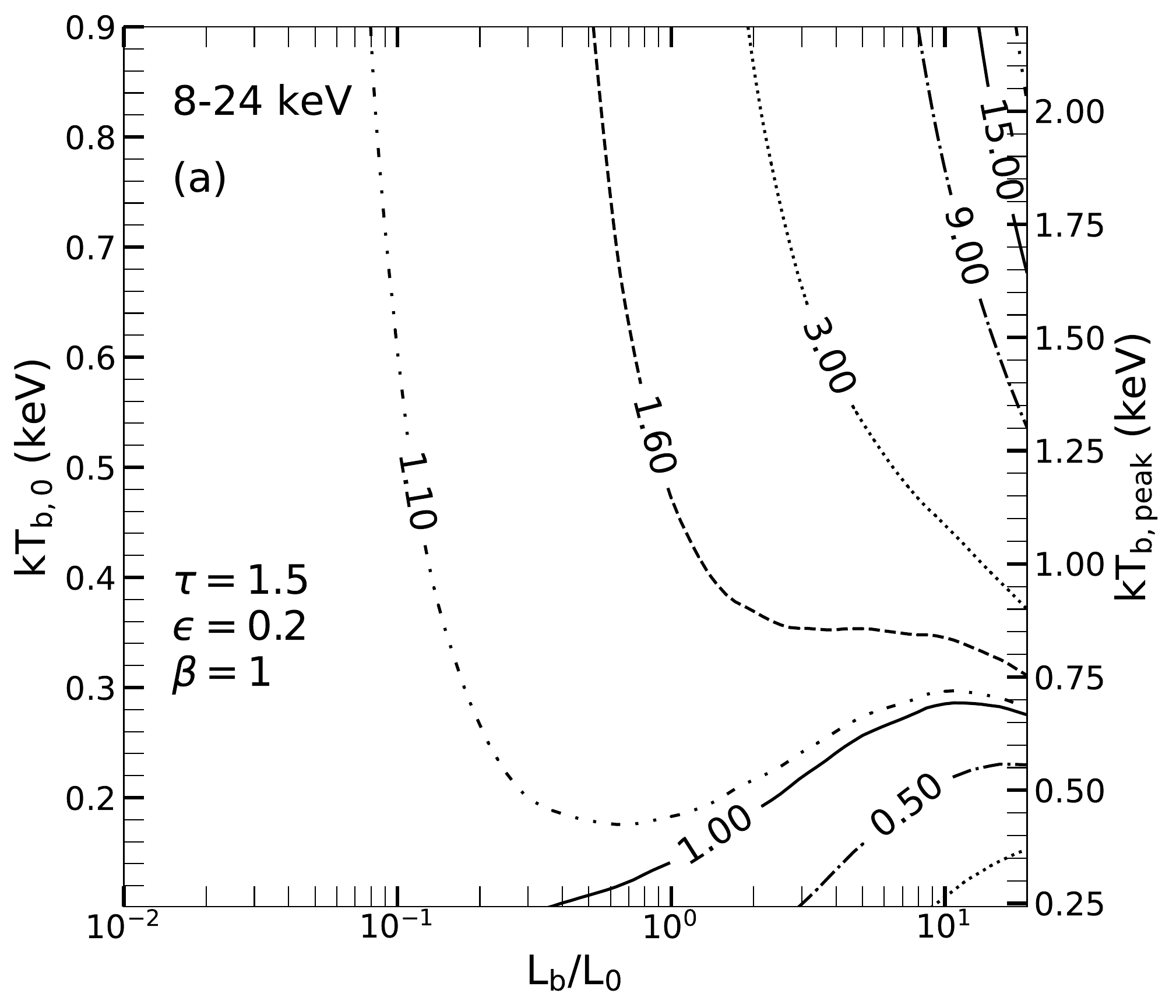}
    \includegraphics[width = 0.45\textwidth]{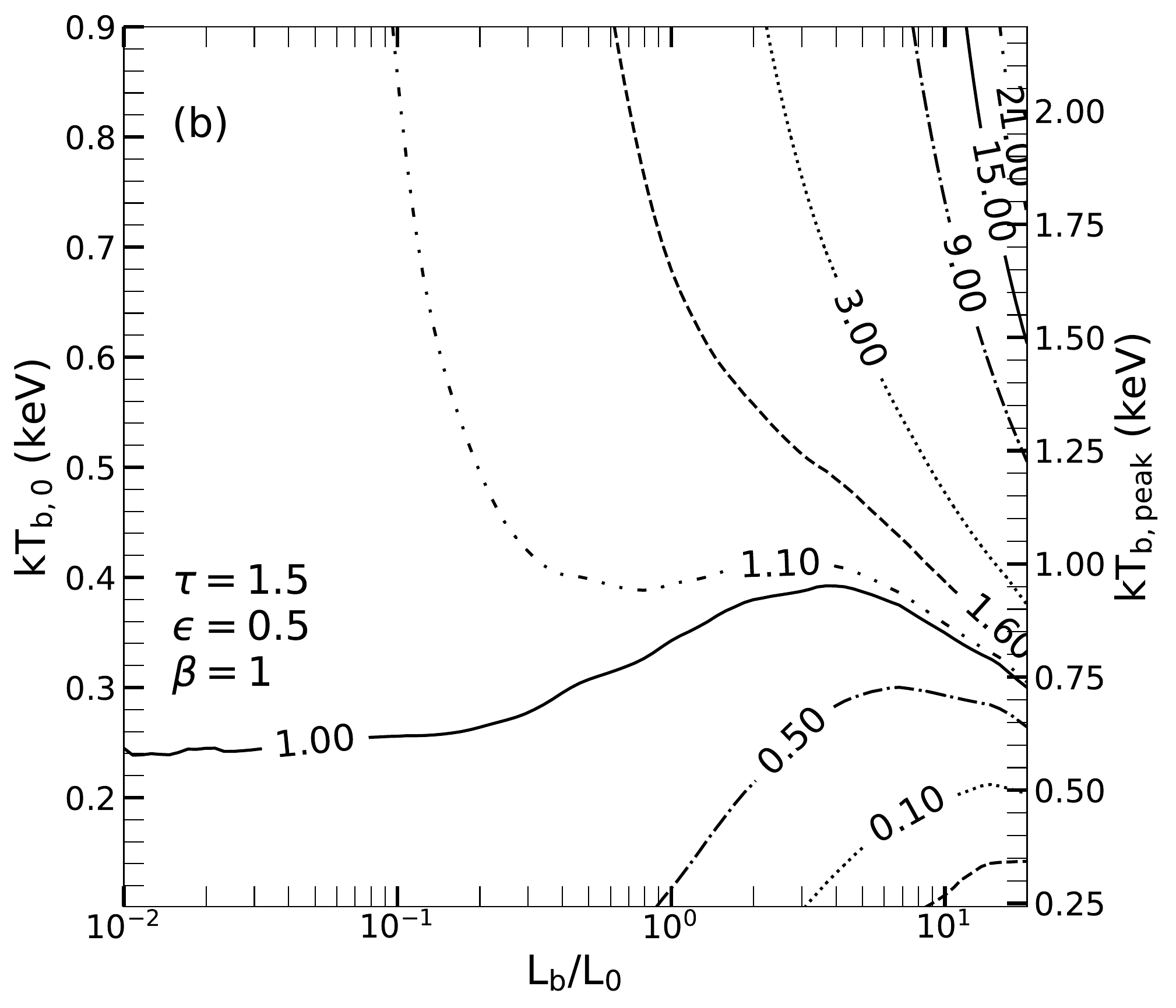}
    \includegraphics[width = 0.45\textwidth]{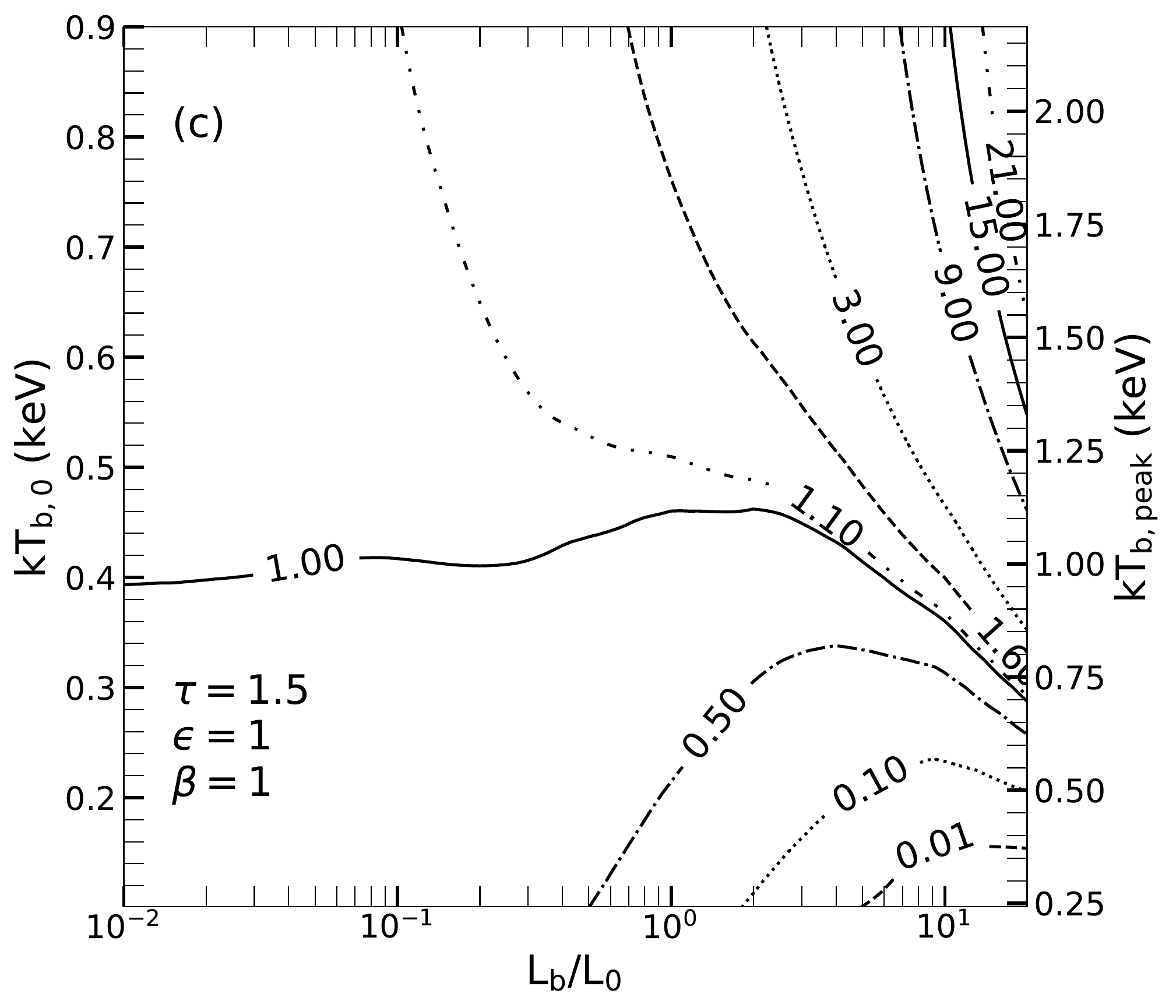}
    \includegraphics[width = 0.45\textwidth]{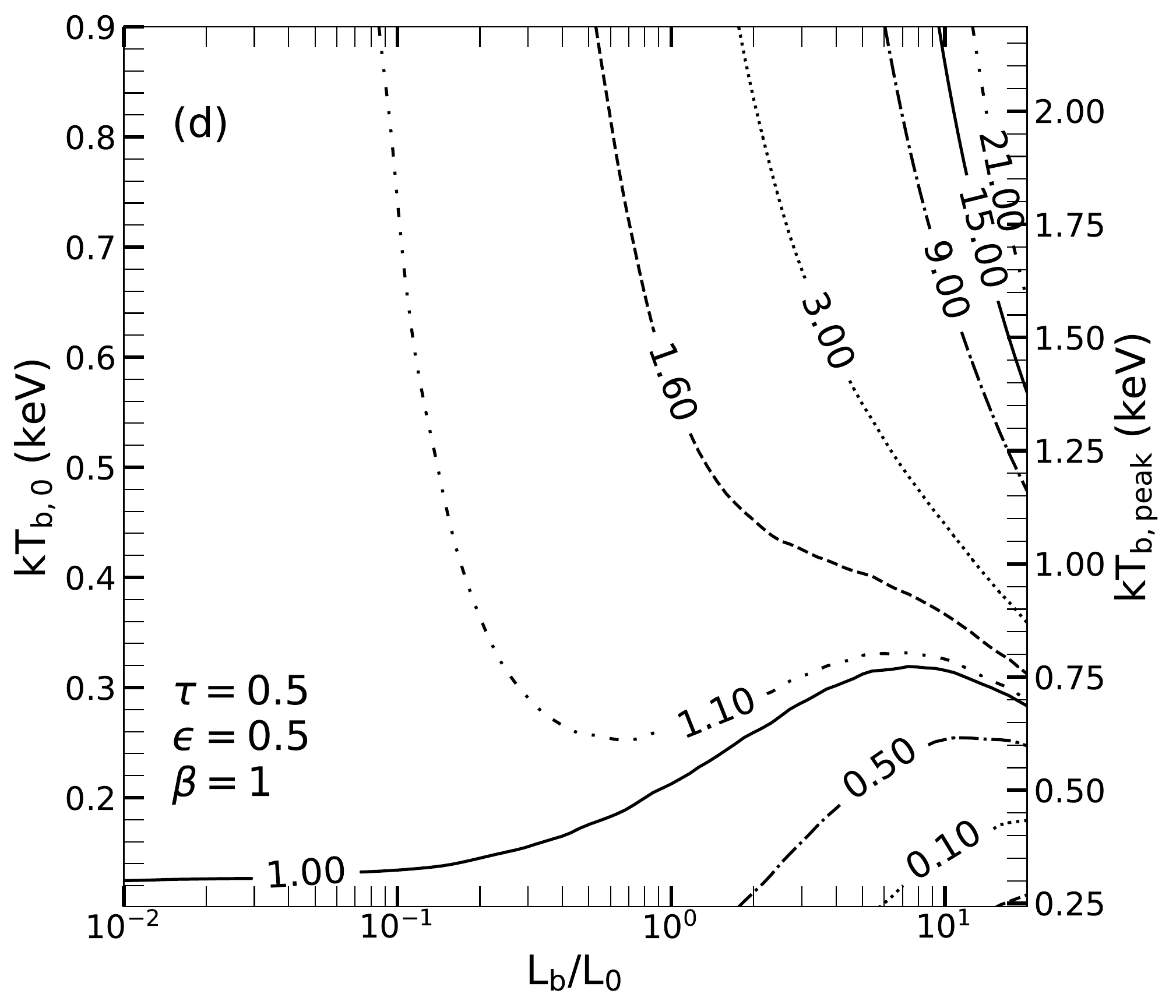}
\caption{Contours of the $8$--$24$~\kev\ flux ratio,
  \frone\ (Eq.\ref{eq:f824}), illustrating the effects of Compton
  cooling of an accretion disc corona due to an X-ray burst. The
  fluxes are computed from the spectra predicted for bursts with
  initial temperature $kT_{\mathrm{b,0}}$ and luminosity
  $L_{\mathrm{b}}/L_0$ and compared to the corresponding model with
  $L_{\mathrm{b}}=0$. All results shown here assume the accretion
  luminosity is constant throughout the burst (i.e., $\beta=1$;
  Eq.~\ref{eq:beta}). The contour levels are the same in all panels,
  with \frone$=1$ indicated by the black solid line. Panels (a), (b)
  and (c) show the impact of a larger aspect ratio of the corona,
  $\epsilon$, which increases the fraction of burst photons entering
  the corona (Eqs.~\ref{eq:ls} and~\ref{eq:fns}) and enhances cooling
  (Fig.~\ref{varyTauEpsilonConstantB}). Panel (d) plots contours for a
  lower coronal optical depth $\tau=0.5$, which moderately reduces the
  cooling effects for the same $\epsilon$. All four panels show that
  \frone\ can either increase or decrease during the burst depending
  on the value of $kT_{\mathrm{b,0}}$, $\tau$ and $\epsilon$. The
  right-hand axis on each panel shows the peak blackbody temperature
  reached by the X-ray burst, $kT_{\mathrm{b,peak}}$.}
    \label{fluxRatioMiddle}
\end{figure*}

To quantify the effects of coronal cooling and the link to $\epsilon$
and $\tau$, we compute the following flux ratios in the $8$--$24$ keV and
$30$--$50$ keV band directly from the spectra calculated by \textsc{eqpair}:
\begin{equation}\label{eq:f824}
    \Phi_{\mathrm{8-24\ keV}}=\frac{F_{\mathrm{8-24\ keV}}(L_{\mathrm{b}}/L_0) }{ F_{\mathrm{8-24\ keV,0}}}
\end{equation}
and
\begin{equation}\label{eq:f3050}
    \Phi_{\mathrm{30-50\ keV}}=\frac{F_{\mathrm{30-50\ keV}}(L_{\mathrm{b}}/L_0)}{ F_{\mathrm{30-50\ keV,0}}}.
\end{equation}
The fluxes $F_{\mathrm{8-24\ keV}}(L_{\mathrm{b}}/L_0)$ and
$F_{\mathrm{30-50\ keV}}(L_{\mathrm{b}}/L_0)$ are measured during the
burst while $F_{\mathrm{8-24\ keV,0}}$ and $F_{\mathrm{30-50\ keV,0}}$
are the pre-burst fluxes. The $8$--$24$~keV band is commonly used in
\nustar\ analysis \citep[e.g.,][]{TheNuSTARExtragalacticSurveys}, and,
as seen below, covers an interesting range of energies that is
sensitive to both X-ray
burst and coronal emission. In contrast, the $30$--$50$~keV band
focuses on the impact of the burst on the corona, and can be
investigated by the IBIS/ISGRI instrument on \textit{INTEGRAL}
\citep[e.g.,][]{TheFirstIBIS/ISGRI}. 

Figure~\ref{fluxRatioMiddle} plots contours of \frone as a function of
the burst luminosity $L_\mathrm{b}/L_0$ and initial burst temperature
$kT_\mathrm{b,0}$. Panels a, b and c show the effect of an increasing
coronal aspect ratio, with $\epsilon=0.2$, $0.5$ and $1$,
respectively. In those three panels the coronal optical depth is held
at $\tau=1.5$, while panel d shows the results for a lower coronal
optical depth of $\tau=0.5$ ($\epsilon=0.5$ in this case). The solid
line in each panel indicates \frone$=1$, which distinguishes the regions
with increasing \frone from those with decreasing
\frone. Interestingly, the flux in this energy band may increase
during the burst up to a factor of $\gtrsim20$, particularly when the corona subtends only a small
angle as seen from the NS (i.e., panel (a)) and when the burst has a
larger initial temperature. Note that the corona will still cool in
these cases (e.g., Fig.~\ref{spectraVaryingLbConstantB}), but the
softening of the spectrum is more than offset by the burst spectrum
entering into the energy band. In this situation, the shape of the
contours are most sensitive to the coronal parameters when
$L_{\mathrm{b}}/L_0 \lesssim 1$. However, if $kT_{\mathrm{b,0}}$ is
low enough (less than approximately $0.4$~keV) then a flux deficit will
occur in this band; however, the magnitude of the drop and the
luminosity at which it occurs will depend on the aspect ratio of the
corona. In this case, as the observed spectrum will not be strongly
influenced by the burst spectrum entering the $8$--$24$~\kev\ band, \frone is more sensitive to the value of
$\epsilon$ even if $L_{\mathrm{b}}/L_0 > 1$. Finally, a lower optical depth corona also reduces the region
of parameter space in which \frone$<1$, although the effect is weaker
than changes in the aspect ratio.



%
\begin{figure*}
    \centering
    \includegraphics[width = 0.45\textwidth]{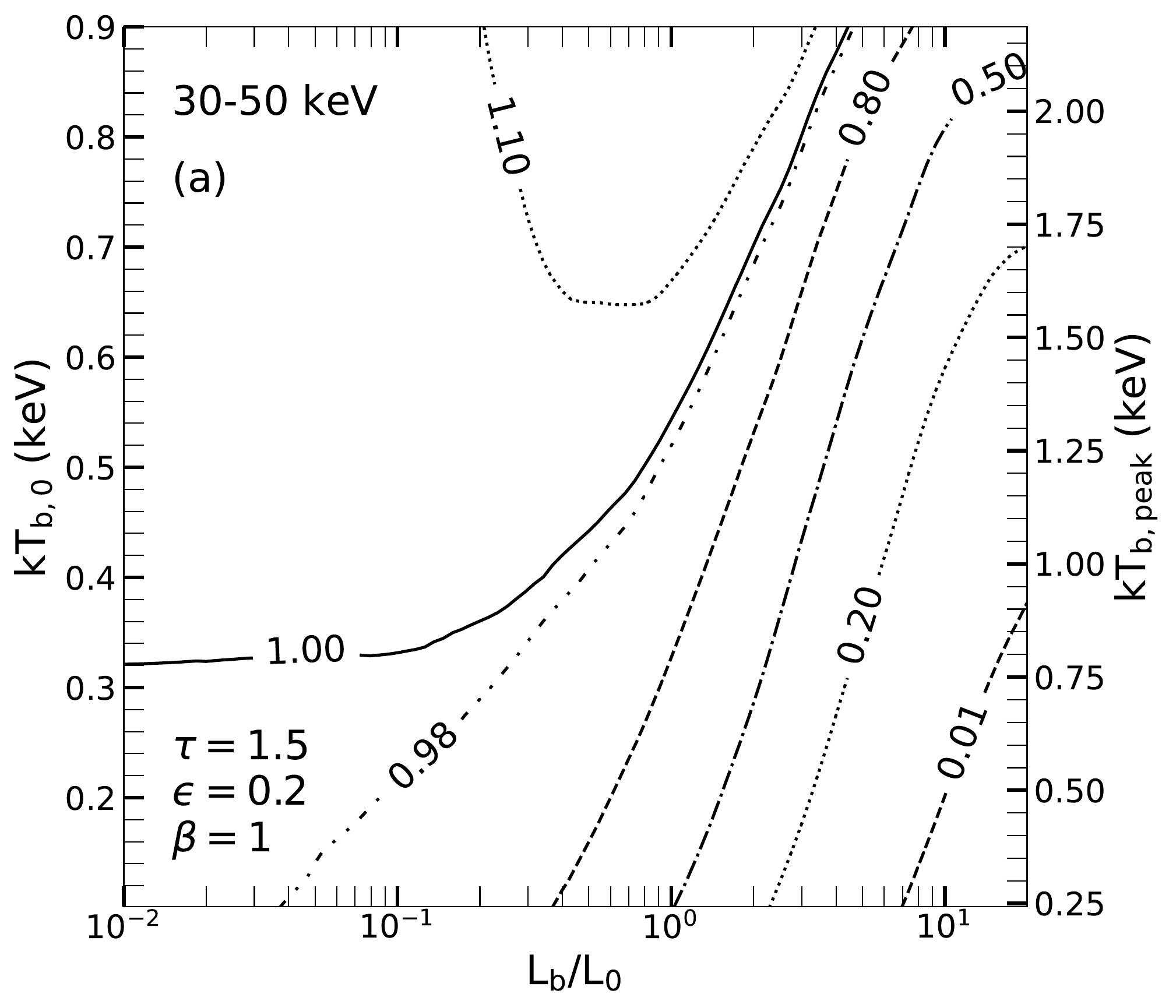}
    \includegraphics[width = 0.45\textwidth]{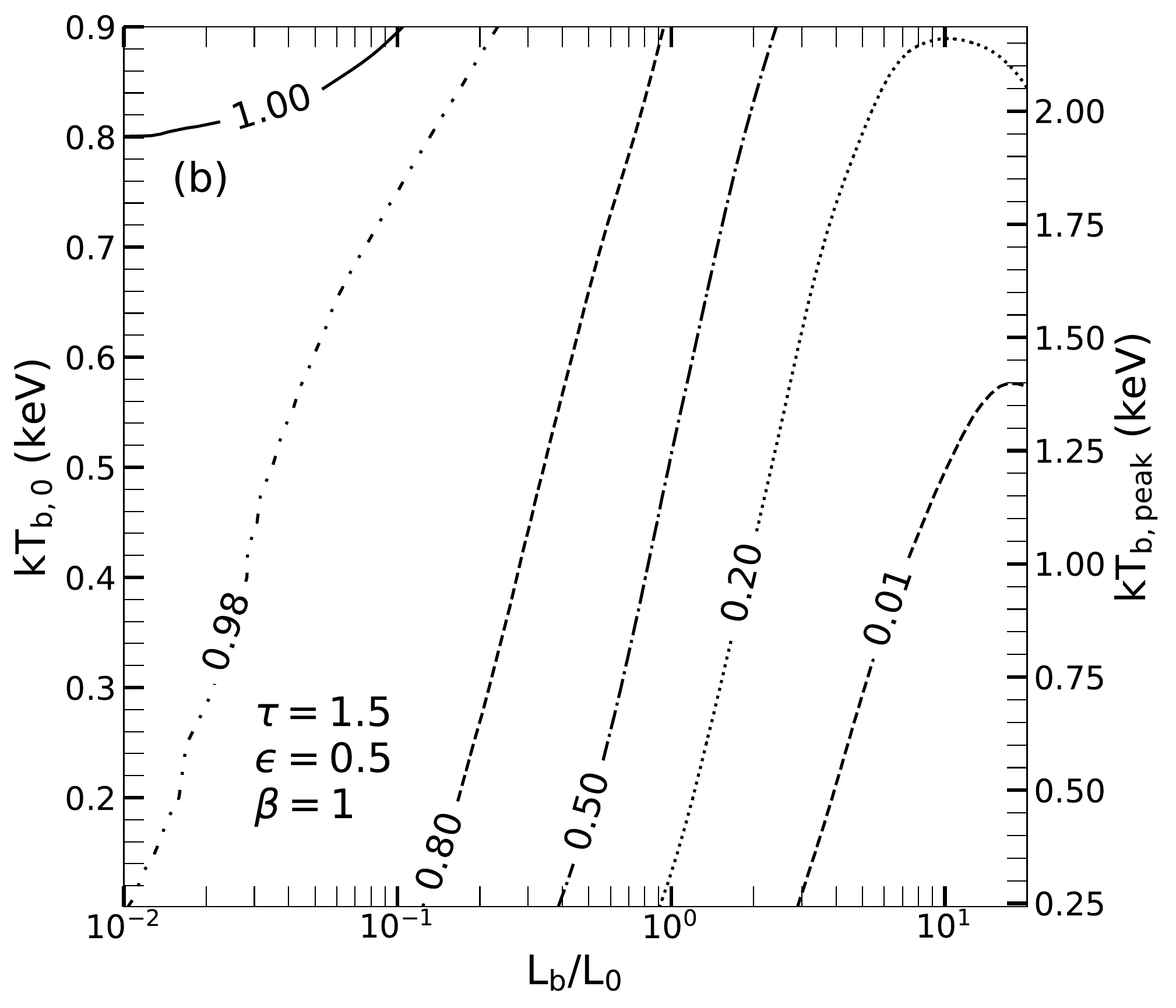}
    \includegraphics[width = 0.45\textwidth]{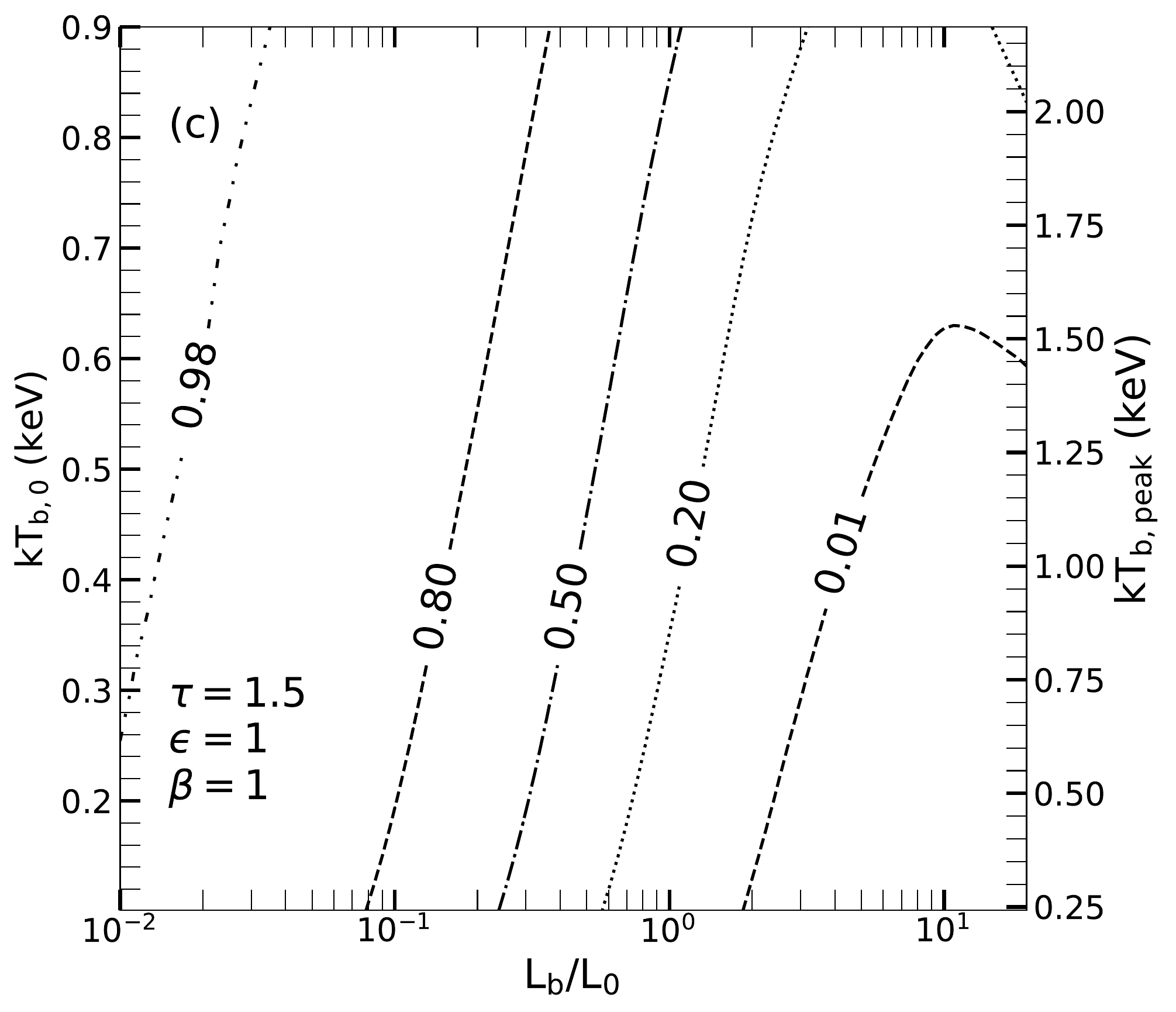}
    \includegraphics[width = 0.45\textwidth]{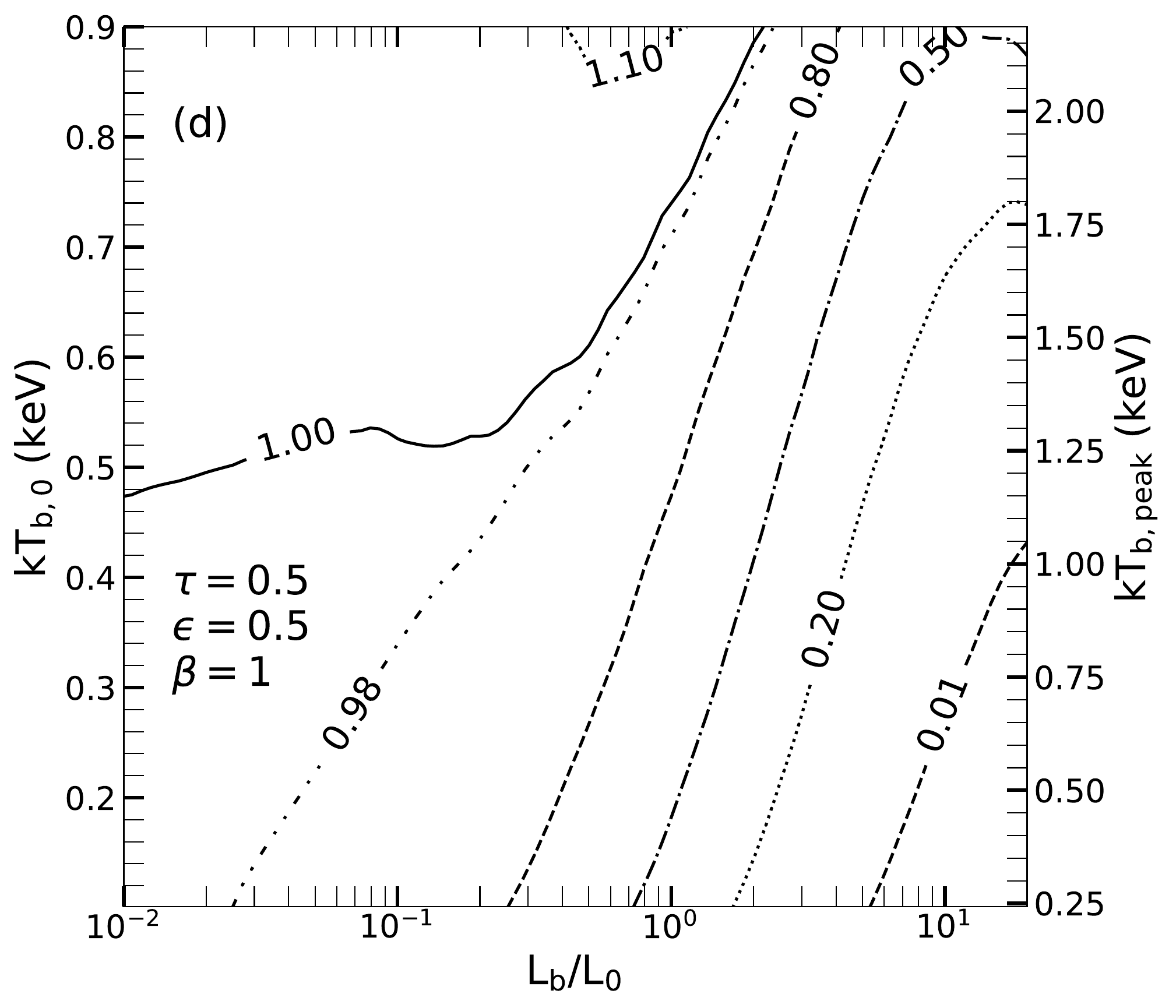}
\caption{As in Fig.~\ref{fluxRatioMiddle}, but now showing contours of
  the $30$--$50$~kev\ flux ratio, \frtwo\ (Eq.~\ref{eq:f3050}). At
  these higher energies, the cooling effects of X-ray bursts
  significantly reduces the emitted flux from the corona, even when
  $L_{\mathrm{b}}/L_0 \sim 0.1$. However, the flux can increase in this
  band by a few percent when $kT_{\mathrm{b},0} \ga 0.5$~\kev\ and
  $L_{\mathrm{b}}/L_0 \la 2$ when the burst interacts only weakly with
  the corona (i.e., $\epsilon=0.2$ or $\tau=0.5$; panels (a) and (d),
  respectively). In these cases, a small part of the burst photosphere
  enters into the band, slightly increasing the flux.}
    \label{fluxRatioHigh}
\end{figure*}

While the X-ray burst can either increase or decrease the emission in
the $8$--$24$ keV band, coronal cooling almost always leads to a
decrease in \frtwo (Eq.~\ref{eq:f3050}) for all $kT_\mathrm{b,0}$
(Figure~\ref{fluxRatioHigh}). However, if the burst interacts only
weakly with the corona, either through a low $\epsilon$ or a small
$\tau$, \frtwo can increase by a few percent for
$kT_\mathrm{b,0}\ga0.5$ keV and $L_\mathrm{b}/L_0\la 2$ because the
small rise in soft emission still outweighs the decrease in hard X-ray
flux. The competition between these two effects causes \frtwo\ to
change very slowly in this region of parameter space with the flux ratio changing only in the 2nd or 3rd
decimal place despite larger changes in $L_\mathrm{b}/L_0$ and
$kT_\mathrm{b,0}$. These slight variations in \frtwo\ lead to small wiggles in the contour lines, most notably when \frtwo$=1$
in Fig.~\ref{fluxRatioHigh}(d). In general, we conclude that observing a small increase in \frtwo would be a clear
indicator of a corona that only weakly interacts with an X-ray
burst. In contrast, Fig.~\ref{fluxRatioHigh}(b) and (c) show that
hard X-ray flux deficits appear for burst luminosities as low as
$L_{\mathrm{b}}/L_0 \sim 0.1$. The amount of change in \frtwo is linked
to the cooling efficiency, which is maximized for low
$kT_\mathrm{b,0}$ and large values of $\tau$ and $\epsilon$. Indeed,
when $L_{\mathrm{b}}/L_0 > 1$, the $30$--$50$~\kev\ flux can fall to
less than 1\% of its pre-burst value when $kT_{\mathrm{b,0}} \lesssim
0.4$~keV. However, higher temperature bursts have a smaller impact in
this energy band and
should therefore be more useful for estimating the coronal properties using the
flux deficit.




\subsection{Impact of an Increasing Accretion Rate During the Burst}
\label{sub:increasebeta}
\begin{figure*}
    \centering
    \includegraphics[width = 0.45\textwidth]{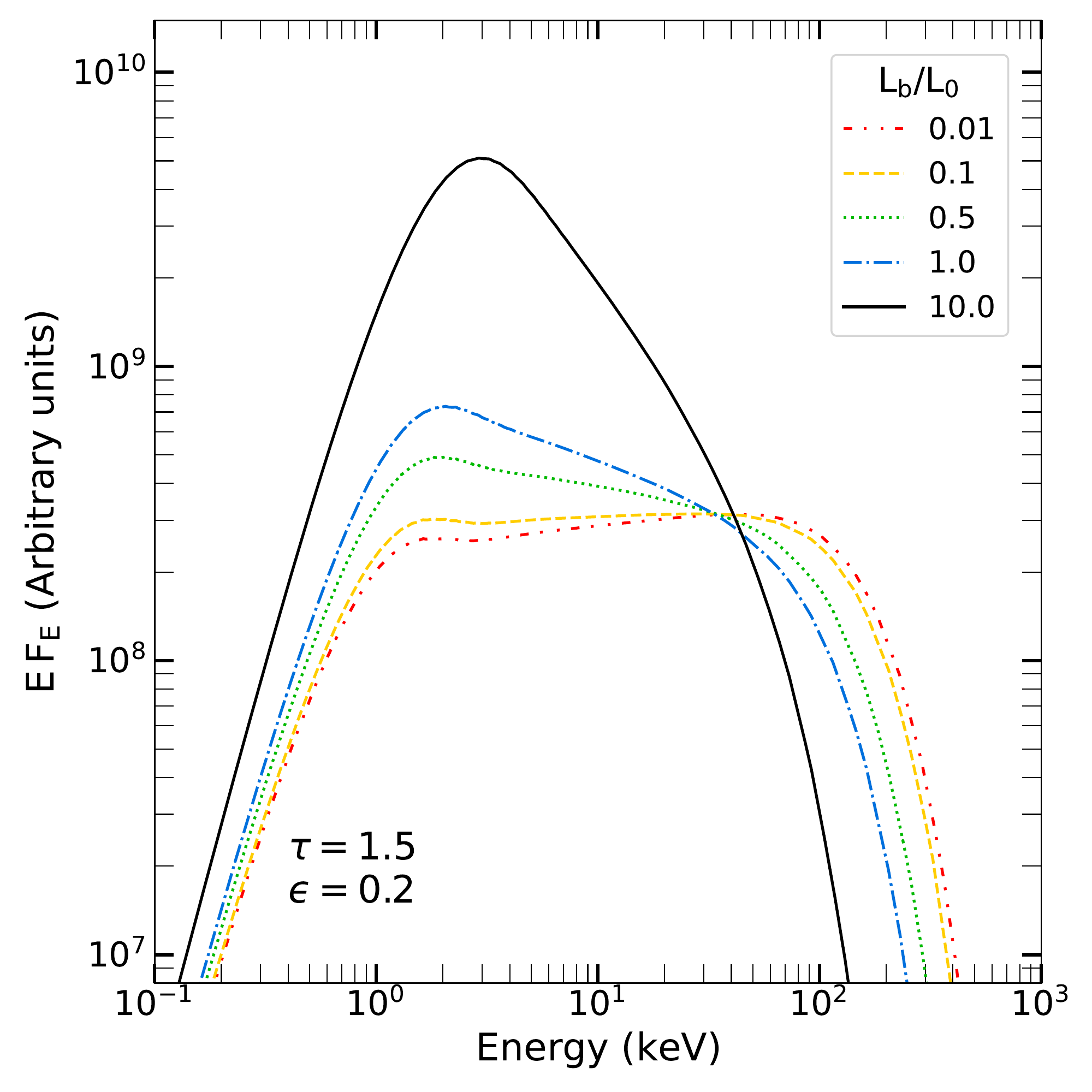}
    \includegraphics[width = 0.5\textwidth]{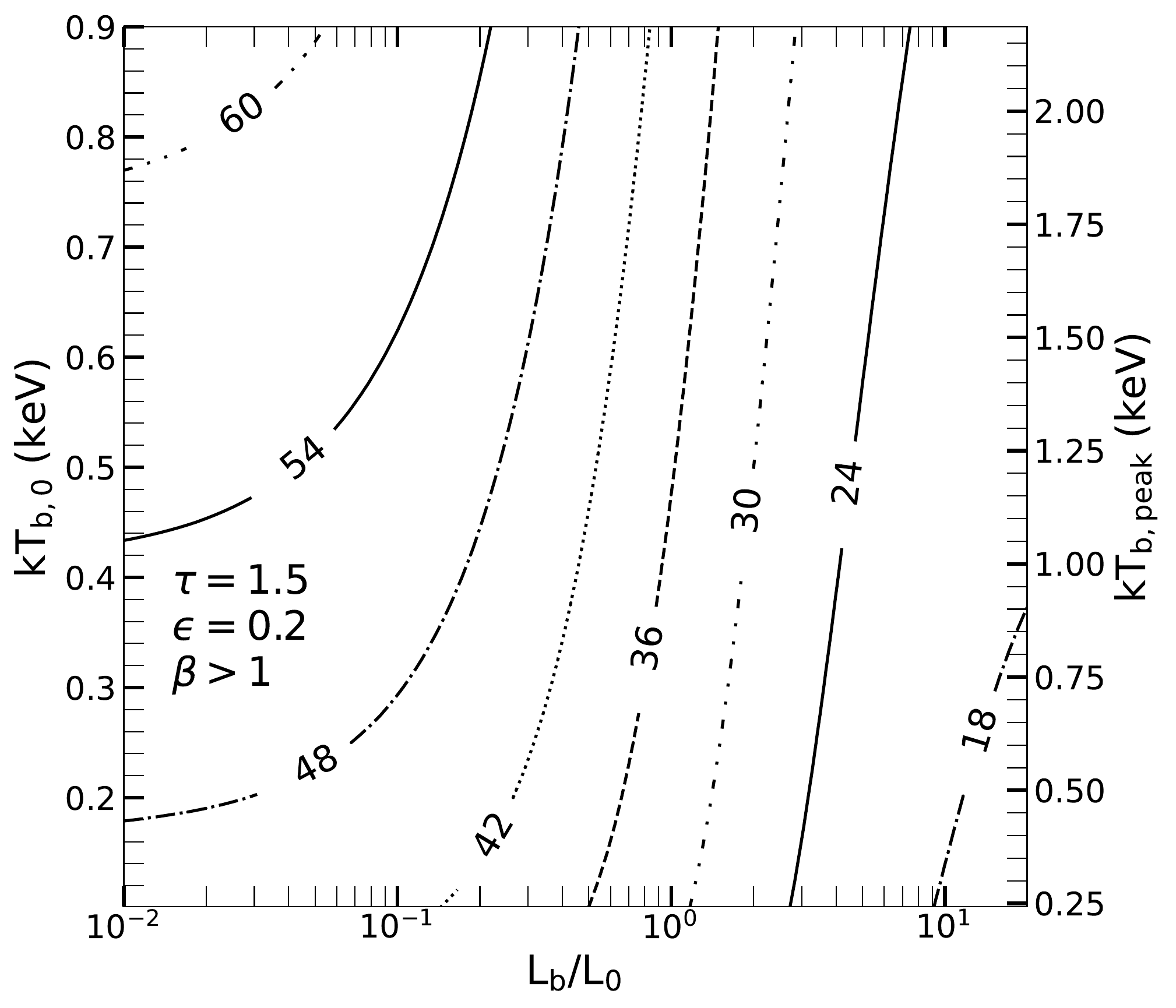}
\caption{As in Figure~\ref{spectraVaryingLbConstantB}, but now showing the effect of increasing the accretion luminosity during the burst (i.e., $\beta > 1$; Eq.~\ref{eq:beta}). 
The non-zero $\beta$ increases the power dissipated in the corona as
the burst luminosity rises (Eq.~\ref{eq:coronaL}). Therefore, the
X-ray burst is unable to cool the corona as effectively in this
scenario.}
    \label{spectraVaryingLbIncreasingB}
\end{figure*}

As mentioned above, numerical simulations of the interaction between an X-ray burst and an accretion disc indicate that Poynting-Robertson
drag may increase the accretion rate onto the NS during the burst
\citep[e.g.,][]{fbb19}. In this section we use Eq.~\ref{eq:beta} to examine how this increase
in accretion power will influence the interaction between the burst
and the corona. An important difference in the energy budget between
the $\beta>1$ and $\beta=1$ cases is the increase in the energy
dissipated in the corona, $L_\mathrm{c}$, which increases linearly
with $\beta$ (Eq.~\ref{eq:coronaL}). However, since only
$L_\mathrm{NS}$ (Eq.~\ref{eq:nsL}), but not $L_\mathrm{b}$, depends on
$\beta$, the total soft luminosity (Eq.~\ref{eq:ls}) increases with
$\beta$ slower than $L_\mathrm{c}$. Hence, the increase in soft
photons injected into the corona cannot completely compensate for the
increase in $L_\mathrm{c}$ due to $\beta > 1$, and the additional
coronal heating partially offsets the Compton cooling driven by the
burst photons (Figure~\ref{spectraVaryingLbIncreasingB}) so that coronal cooling can be mitigated by $\gtrsim60\%$.
In fact, even for burst luminosities as large as  $L_\mathrm{b}/L_0=10$,
a power-law is still in the resulting spectrum. The reduced Compton
cooling is seen in the equilibrium coronal electron temperatures
(compare the right panels of Fig.~\ref{spectraVaryingLbConstantB} and
Fig.~\ref{spectraVaryingLbIncreasingB}), and will have a
noticeable impact on the detected hard X-ray flux.

\begin{figure*}
    \centering
        \includegraphics[width = 0.45\textwidth]{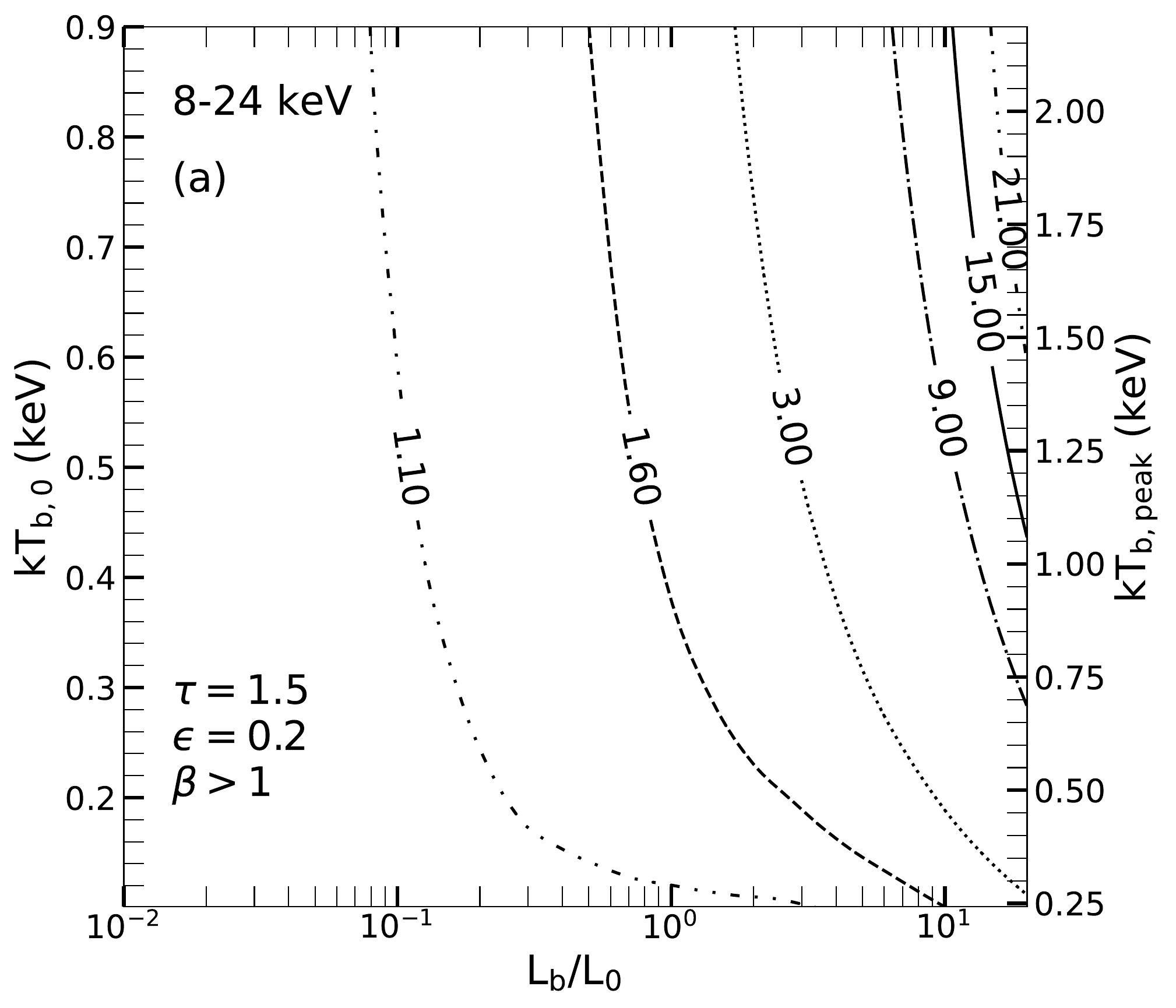}
    \includegraphics[width = 0.45\textwidth]{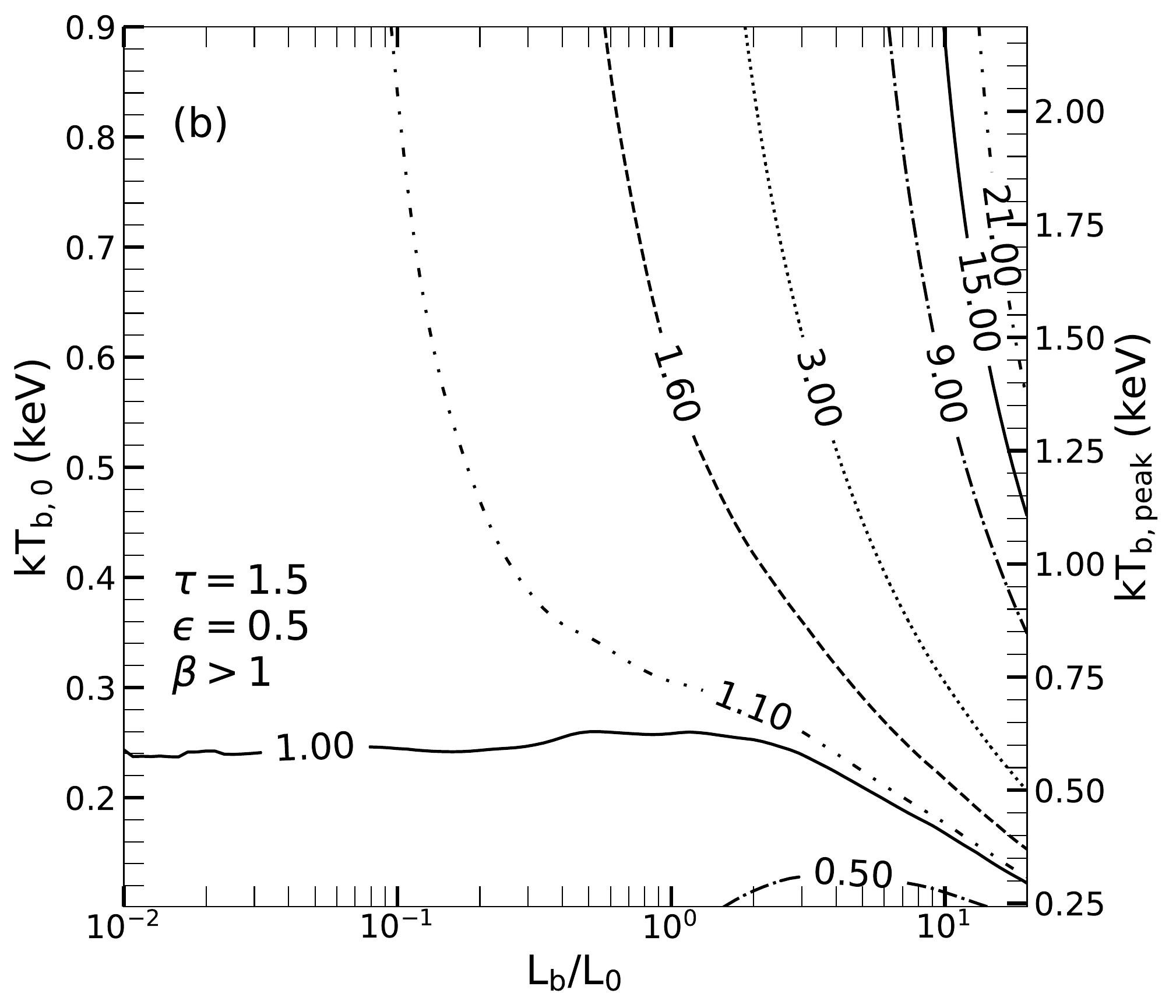}
    \includegraphics[width = 0.45\textwidth]{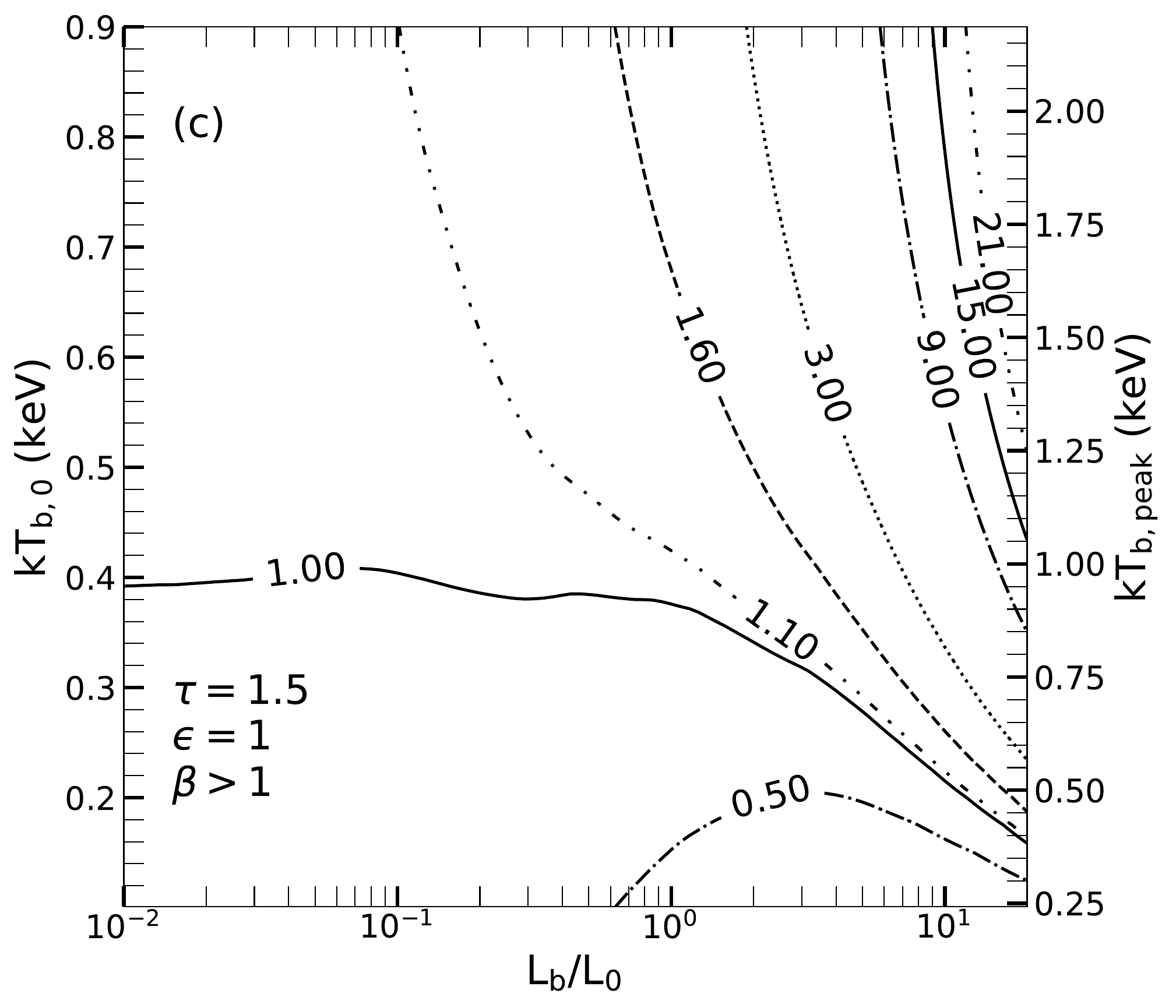}
    \includegraphics[width = 0.45\textwidth]{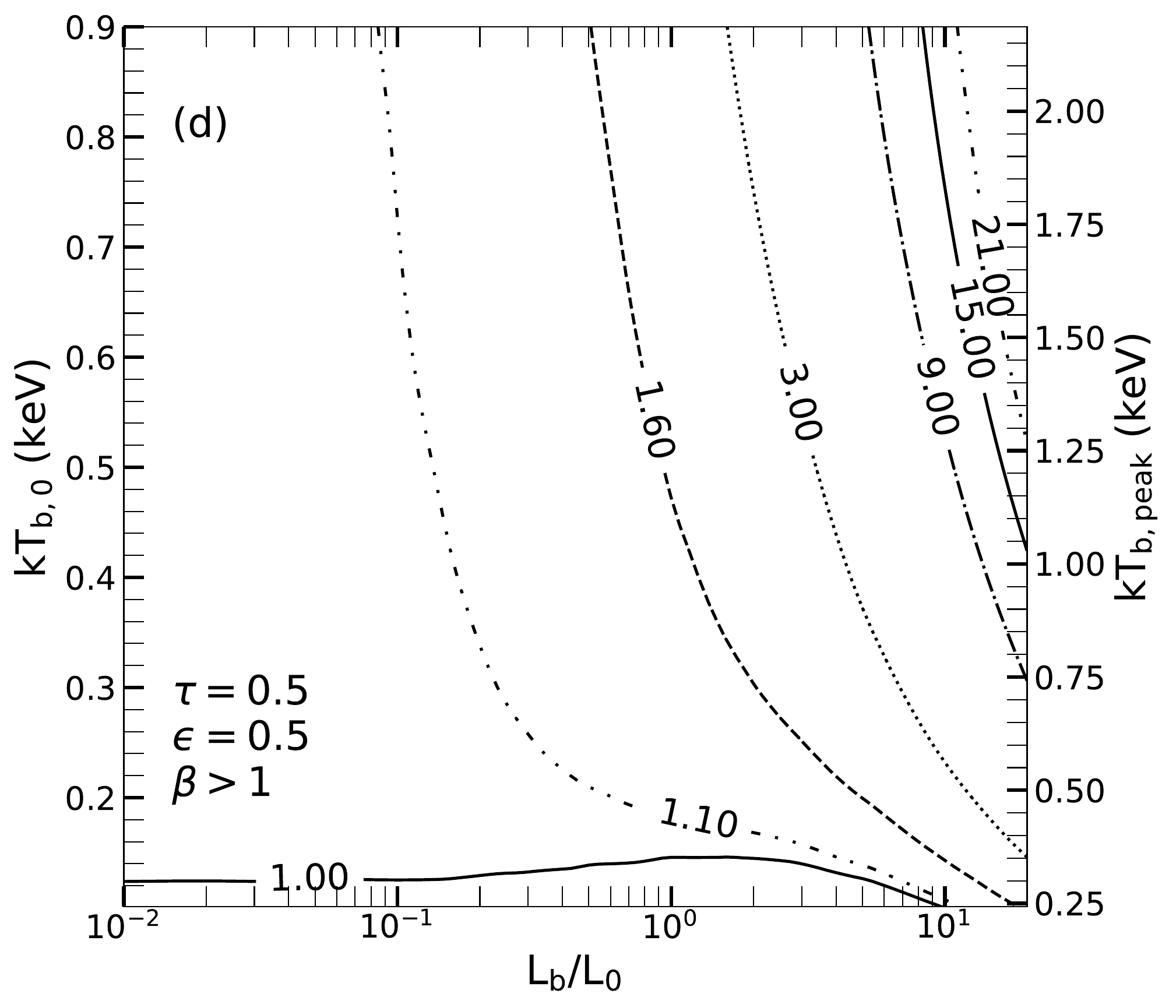}
\caption{As in Figure~\ref{fluxRatioMiddle}, but now for the situation
  where the accretion luminosity increases during the burst (i.e.,
  $\beta > 1$; Eq.~\ref{eq:beta}). The larger $L_c$ during the burst
  reduces the effectiveness of the Compton cooling and the spectrum
  retains a Comptonized tail even when $L_{\mathrm{b}}/L_0 \ga 1$
  (Fig.~\ref{spectraVaryingLbIncreasingB} (Left)). Thus, \frone$\ga 1$
  for nearly all burst luminosities except when $\epsilon=1$ and a
  large fraction of the burst passes through the corona (panel (c)).}
    \label{fluxIncMiddle}
\end{figure*}

Figure~\ref{fluxIncMiddle} shows the impact of the extra energy
dissipated in the corona when $\beta>1$ in the $8$--$24$ keV band flux
ratios, \frone (Eq.~\ref{eq:f824}). The contour levels shown here are
the same as in Fig.~\ref{fluxRatioMiddle} which showed the results for
the $\beta=1$ case. The increases in \frone are largely unchanged from
the previous scenario, as these are driven by the shape of the burst
spectrum which is not impacted by $\beta$. However, the region of
parameter space where \frone$<1$ is smaller when $\beta > 1$ and the
flux ratios fall to only $\sim 0.5$, several times larger than in the
$\beta=1$ case. In fact, when only a small fraction of the burst interacts
with the corona ($\epsilon=0.2$; panel a), \frone$>1$ everywhere in
the computational domain. Therefore, if the accretion rate does
increase during an X-ray burst then only a moderate decrease in the
$8$--$24$~\kev\ flux might be observed.



%
\begin{figure*}
    \centering
    \includegraphics[width = 0.45\textwidth]{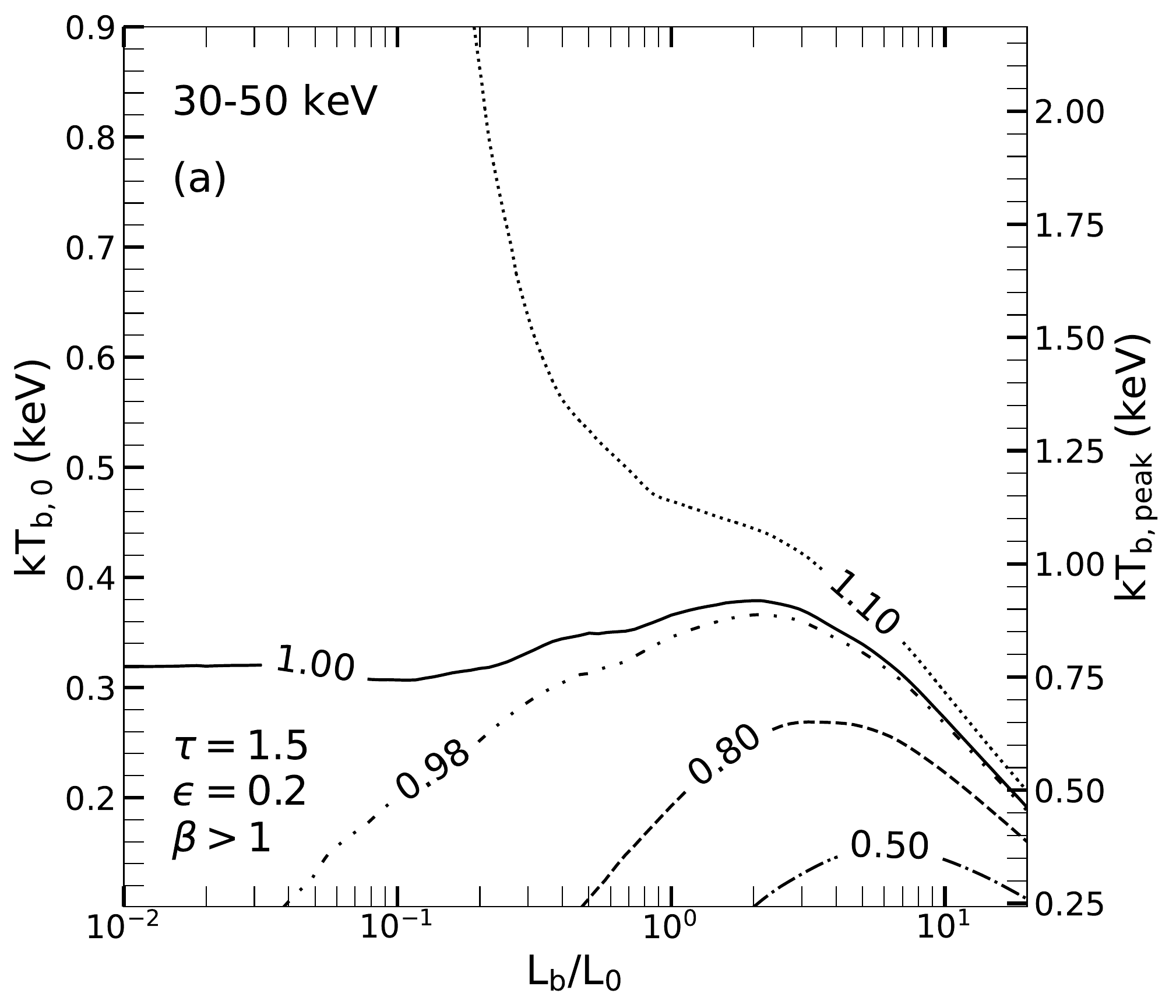}
    \includegraphics[width = 0.45\textwidth]{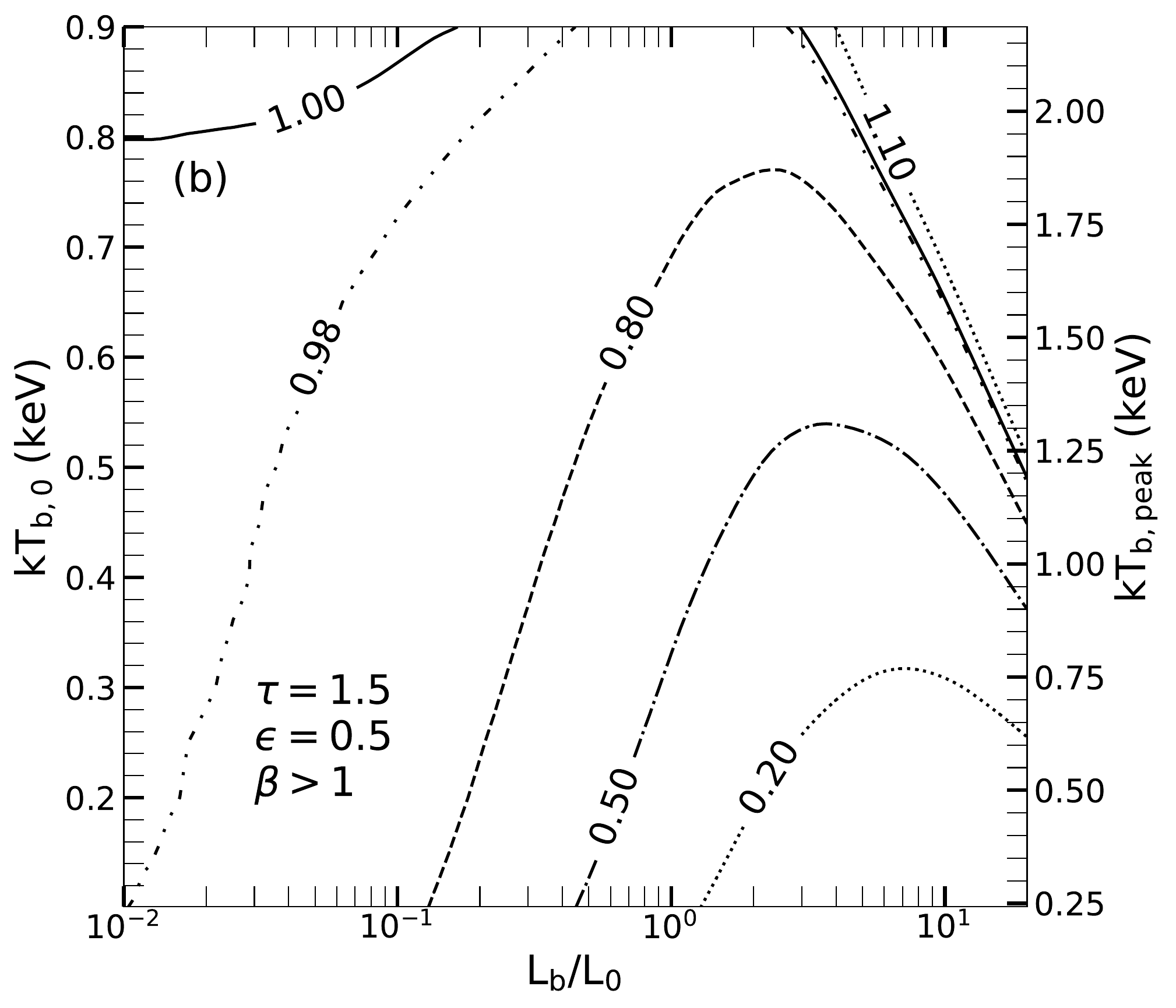}
    \includegraphics[width = 0.45\textwidth]{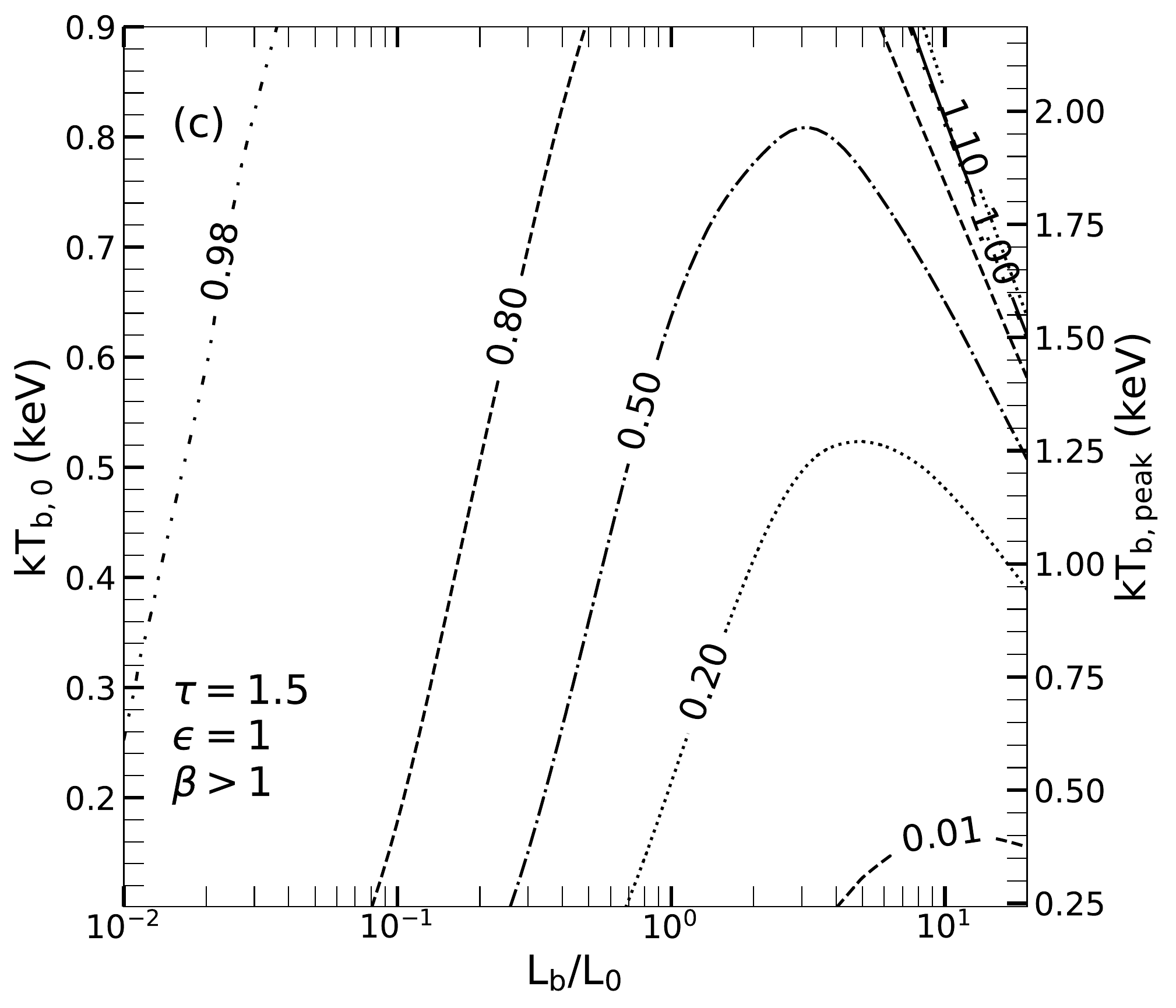}
    \includegraphics[width = 0.45\textwidth]{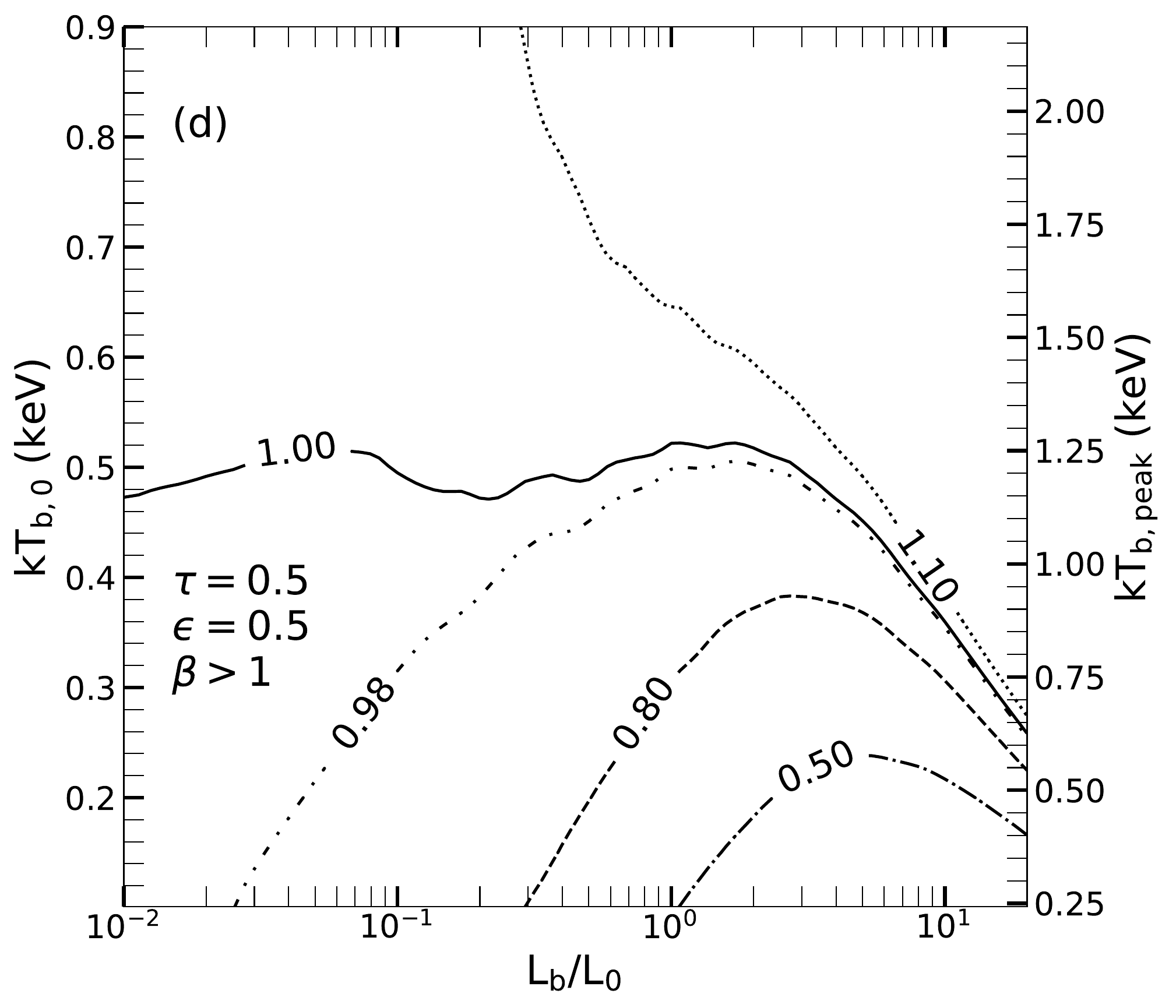}
\caption{As in Figure~\ref{fluxRatioHigh}, but now we let the accretion
  luminosity increase as a function of $L_{\mathrm{b}}/L_0$ (i.e.,
  $\beta > 1$; Eq.~\ref{eq:beta}). Although Compton cooling is less
  effective during the burst, a significant flux deficit (i.e.,
  \frtwo$< 1$) is predicted whenever the burst can efficiently
  interact with the corona (panels (b) and (c)). However, as in the
  $\beta=1$ case, \frtwo\ can increase by a few percent for bursts
  that do not scatter strongly in the corona (panels (a) and (d)).}
    \label{fluxIncHigh}
\end{figure*}

The change in cooling efficiency if $\beta>1$ leads to more drastic
differences in the $30$--$50$ keV band as shown in
Figure~\ref{fluxIncHigh}. The region where \frtwo$>1$
significantly increases when the burst interacts weakly with the
corona (panels (a) and (d)). Similar to the $\beta=1$ case
(Fig.~\ref{fluxRatioHigh}), \frtwo$\approx 1$ over a wide range of
parameter space in these two panels, leading to wiggles
appearing in the \frtwo$=1$ contours. In addition, there is a small region (in
the upper-right corner of panels (b) and (c)) where \frtwo$>1$ even
when the corona has a large aspect ratio. This increase of a few
percent in the $30$--$50$~\kev\ flux is attributed to the increase in
$L_\mathrm{c}$ (Eq.~\ref{eq:coronaL}), which generates a non vanishing
high-energy electron contribution (e.g.,
Fig.~\ref{spectraVaryingLbIncreasingB}) that produces a spectrum that extends into the
$30$--$50$ keV band for all burst and coronal properties we
tested. The high-energy electron contribution increases with
$L_\mathrm{c}$, and, combined with soft photons scattered to higher
energies as the burst luminosity increases (Eq.~\ref{eq:TBB_b}), can
cause \frtwo to increase during the burst. Similar to the
$8$--$24$~\kev\ flux ratio, the drop in \frtwo\ is reduced when $\beta
> 1$ due to the increase in the coronal dissipation as the burst
luminosity increases. However, large hard X-ray flux deficits are
still expected in this band unless the burst interacts weakly with the
corona.

\section{Discussion}
\label{sect:discuss}


The results presented above have shown that an X-ray burst can
significantly cool the accretion disc corona which will therefore substantially
alter the shape of the hard X-ray spectrum. In addition to the
strength and temperature of the burst spectrum, the change in the hard
X-ray flux depends on both the optical depth and aspect ratio of the
corona. The $8$--$24$~\kev\ flux may either increase or decrease during the
burst, but the $30$--$50$~\kev\ flux will almost always decrease. The
hard X-ray deficits can be mitigated if the X-ray burst induces an
increase in the local accretion rate. Thus, examining the behavior of
the hard X-ray flux during the burst may provide information on the
coronal properties of the accreting system.

\subsection{Comparison to Observations}
\label{sub:comparisonObservations}
As mentioned in Sect.~\ref{sect:intro}, there have been multiple claims
of hard X-ray shortages during X-ray bursts discussed in the
literature. Here, we compare these observations to the results presented above. Given the idealized nature of the model, only qualitative conclusions
can be drawn by comparing to observations. We focus on the calculations in the $30$--$50$
keV band, as it more strongly reflects the impact of coronal
cooling. However, we would expect a band extending to energies
$>50$~keV to be more impacted by coronal cooling, and a band extending
to $<30$~keV to be impacted less.

\citet{Maccarone2003} reported the first hard X-ray shortage during an
X-ray burst from Aql~X-1, with the flux in the $30$--$60$~keV band
dropping to $\sim50\%$ of its pre-burst value. Strong Compton cooling of the corona during bursts from Aql~X-1 was also inferred
  by \citet{Chen2013} from stacking multiple \rxte\ burst light
  curves. At the point of the burst \citet{Maccarone2003} determined with a Comptonization model the coronal optical depth to be $\tau \sim 4.2$. The
authors also found that the emission in the $15$--$30$~keV band
increased by $\sim20\%$ during the burst which may indicate a large
burst temperature and luminosity (e.g.,
Fig.~\ref{fluxRatioMiddle}). In this case, our results in the
$30$--$50$~\kev\ band (Figs.~\ref{fluxRatioHigh}
and~\ref{fluxIncHigh}) are consistent with a drop of about of factor of
$2$, especially if $\epsilon > 0.5$. However, the large optical depth
measured by \citet{Maccarone2003} suggests a high cooling efficiency,
which indicates that an enhanced accretion rate during the burst may
have mitigated the drop in the hard X-ray emission (e.g.,
Fig.~\ref{fluxIncHigh}). Our results seems to imply that at the time
of this burst the corona of Aql~X-1 had a large aspect ratio, and that
the burst enhanced the accretion power dissipated within the corona.


A deficit of $\sim 50\%$ in the $30$--$50$~\kev\ band was also
described by \citet{ChenEtAl2012} after stacking multiple \rxte\
lightcurves of bursts from IGR J17473–2721. The maximum deficit
occurred at the peak of the bursts, which may indicate weak interaction
with the corona, or an enhanced accretion rate during the
bursts. Interestingly, \citet{ChenEtAl2012} only find the hard X-ray
shortage before the system entered into the high/soft state, and do not find any hard X-ray
deficit for bursts after the system transitioned back into the
low/hard state (their 'lagging low-hard-state'). This may indicate
a change in the coronal structure after the state transition, and
implies that careful analysis of future hard X-ray shortages could
elucidate the evolution of the corona as systems undergo major changes
in their accretion properties.

In another stacking analysis of \rxte\ lightcurves from GS~1826–238, \citet{ji2014}
found a similar factor of two drop in the $30$--$50$ keV band. Using
previous work on the persistent spectrum
\citep[e.g.,][]{Thompson2005,Thompson2008,Cocchi2010}, the coronal
temperature was estimated to fall from $\sim20$~keV to $\sim10$~keV
assuming a fixed $\tau\approx2.6$. These low temperatures imply a
large $\epsilon$, but the drop of only a factor of $2$, on average,
during an X-ray burst, suggests an increase in accretion power that
can help sustain the coronal temperature (e.g.,
Fig.~\ref{fluxIncHigh}), in agreement with the conclusion of \citet{ji2014}.

A dramatic example of coronal cooling was identified by
\citet{Kajava2017} who stacked 123 X-ray bursts detected by
\integral\ from 4U~1728–34. The stacked emission in the $40$--$80$~keV
band drops to a third of the pre-burst emission, and the coronal
temperature is estimated to drop from $\sim21$ keV to $\sim 3.5$
keV. The mean blackbody temperature of the burst is $2.6$~\kev\ placing
it off the top of the panels in Figs.~\ref{fluxRatioHigh}
and~\ref{fluxIncHigh}. The strong drop in flux suggests a high
cooling efficiency and thus a high optical depth and/or a high aspect
ratio, with a minimal increase in accretion rate during the burst. Due
to the great amount of cooling \citet{Kajava2017} speculated if the
drop in temperature could prompt the corona to collapse, which has
since been shown by
\citet{SimulatingtheCollapseofaThickAccretionDisk}. These results
may indicate that the inner accretion flow in 4U~1728-34 is dominated
by a hot, geometrically thick corona.

Recently, \citet{Sanchez2020} stacked \integral\ and \xmm\ burst
spectra from GS~1826–238 and found that the average emission in the
$35$--$70$~keV band dropped by $\sim80\%$ during the bursts while the
emission in the $18$--$35$~keV band showed an
increase. \citet{Sanchez2020} do not report an optical depth, but the
peak blackbody temperature in the stacked spectra is $\sim 2$~\kev,
placing it near the top of Figs.~\ref{fluxRatioHigh}
and~\ref{fluxIncHigh}. This is consistent with the bursts being
detectable in the $18$--$35$~\kev\ band, as seen in our predictions
of \frone (e.g., Fig.~\ref{fluxRatioMiddle}). The large drop in the
$35$--$70$~keV band implies a high cooling efficiency and that the
bursts strongly interacted with the corona (e.g., a large $\epsilon$ and
$\tau$), but did not induce a large increase in the accretion rate. This
is consistent with the conclusions of \citet{Sanchez2020} who find
that the total persistent flux changes by only $\sim10\%$ during the
burst, although the normalization factor of the persistent emission
rises by $30$--$80\%$. Thus, the average bursts analyzed in this system
seem to appear to not have a large impact on the accretion rate.

It is worth emphasizing that it is challenging to infer the magnitude of any change in
the accretion rate through spectral analysis. While
\citet{Sanchez2020} found that the normalization factor for the
persistent emission rises by $30$--$80\%$, other analyses show increases
of $\gtrsim 2$ \citep[e.g.,][]{in'tZand2013,Worpel2015,DegenaarEtAl2016,KeekEtAl2018}. As \citet{Sanchez2020}
notes, these larger increases in accretion rate could be due to only
using the normalization factor to track the evolution of the
persistent spectrum, or because the difference in persistent emission
increase is model dependent. In addition, numerical simulations by
\citet{SimulatingtheCollapseofaThickAccretionDisk} and \citet{fbb19}
both predict a burst-driven enhanced accretion rate of a factor of a
few, and this increase appears to be a natural outcome of the
interaction of the burst with its surroundings. However, the
simulations are currently unable to predict how this enhanced
accretion rate will manifest in the emitted spectrum. Clearly,
additional work is needed in determining the observational
consequences of a burst-driven increase in the accretion rate.


Finally, our results are also consistent with X-ray burst
observations where no hard X-ray shortages were observed during a
burst
\citep[e.g.,][]{in'tZand1999,ChenEtAl2012,jiEtAl2014KS1731-260,DegenaarEtAl2016}. Figs.~\ref{fluxRatioMiddle}--\ref{fluxIncHigh}
show that coronal cooling will occur without a hard X-ray emission
drop, in particular when the cooling efficiency is low (e.g., low
$\tau$ or $\epsilon$),  and/or there is a significant increase in
the coronal heating during the burst. Thus, X-ray bursts that do not
cause a hard
X-ray shortage in the persistent spectrum may still be cooling the
accretion disc corona. In these cases, broadband spectral modeling of
the persistent spectrum during the burst is needed to uncover the
coronal parameters.




\subsection{The Impact of Reflection from the Accretion Disc}
\label{sub:reflection}


The calculations presented in Sections~\ref{sect:calc}
and~\ref{sect:results} are focused on the interaction of the corona
and the burst, and omit the effects of the optically thick accretion
disc which will re-process (or reflect) radiation from both the corona
and the burst. This reflected radiation will be another source of soft
photons \citep[e.g.,][]{Ballantyne2004}, and can contribute to coronal
cooling if a significant fraction enters the corona. It is therefore
worthwhile to consider how the additional soft photons produced by
reflection may impact the predicted \frone and \frtwo. For simplicity
we will only consider the case of a constant accretion rate (i.e. $\beta=1$).

We include the impact of reflection by following
\citet{DegenaarEtAl2018} and define a reprocessed luminosity that
is the sum of the soft luminosity that originated from the neutron star and traversed the corona and the coronal luminosity
\begin{equation}\label{eq:lrep}
L_{\mathrm{rep}} = \alpha\left[L_\mathrm{s}\left(1-e^{-\tau}\right)+L_\mathrm{c}\right],
\end{equation}
where $\alpha=0.1$ is the fraction of the disc that is illuminated, consistent with a scenario of a truncated
accretion disc \citep[e.g.,][]{Barret2000,Zdziarski2003}.
As $L_{\mathrm{rep}}$ originates from the disc, only a fraction
$f_\mathrm{d}$ illuminates the corona as soft luminosity, and the new
pre-burst soft luminosity $L_\mathrm{s,0}$ is
\begin{equation}\label{eq:ls0_new}
L_\mathrm{s,0}=\left \{  \frac{f_\mathrm{d}+f_\mathrm{NS}+f\left [f_\mathrm{NS} f_\mathrm{c}-f_\mathrm{d}\left ( 1-\alpha+\alpha f_\mathrm{c} \right )  \right ]}{1-\alpha f_\mathrm{d}\left ( 1-e^{-\tau} \right )}  \right \}\frac{L}{2}.
\end{equation}
During the burst the soft luminosity is now written as $L_\mathrm{s}$
\begin{equation}\label{eq:ls_new}
L_\mathrm{s}=L_\mathrm{s,0}+\left [ \frac{f_\mathrm{NS}}{1-\alpha f_\mathrm{d}\left (1-e^{-\tau}  \right )} \right ] L_\mathrm{b}.
\end{equation}
The blackbody temperature of the burst (Eq.~\ref{eq:TBB_b})
remains unchanged and is used as the seed photon temperature in
\textsc{eqpair}. The reflection spectrum is included in the
\textsc{eqpair} calculation with an amplitude of $0.2$ \citep{DegenaarEtAl2018}.


From the above equations we see that the cooling effect of
$L_\mathrm{rep}$ on the corona will increase with $L_\mathrm{b}$ and
$\epsilon$ (Eq.~\ref{eq:ls_new}), thus we expect a larger flux deficit
in the hard X-ray band under these conditions. To quantify the effect
of reflection we define the fractional change in the calculated flux
ratios as
\begin{equation}\label{eq:fractionalChange}
\Delta\Phi/\Phi=\left | 1-\frac{\Phi_{\alpha\neq 0} }{\Phi_{\alpha= 0}} \right |,
\end{equation}
where $\Phi_{\alpha\neq 0}$ is the flux ratio with and $\Phi_{\alpha=
  0}$ without reflection. We calculated \dPhi over the same
$L_\mathrm{b}/L_0$ and $kT_\mathrm{b,0}$ ranges and choices of $\tau$
and $\epsilon$ as considered in Sect.~\ref{sect:results}. We find that
\dPhi in the $30$--$50$~keV band is greater than in the $8$--$24$~keV
band since the impact of cooling is more pronounced at higher
energies. As anticipated, the greatest fractional change in the
$30$--$50$~keV band occurs when in geometries where Compton scattering
occurs frequently (e.g., $\epsilon=1$), and when the radiation field
is most effective in removing energy from the corona (e.g.,
$kT_\mathrm{b,0} \lesssim 0.5$ keV and
$L_\mathrm{b}/L_0\gtrsim10$). The magnitude of the change in
flux ratio due to reflection is relatively small with \dPhi in the
$30$--$50$~keV band $\lesssim 10\%$ and \dPhi in the $8$--$24$~keV
band below $5\%$. These results will change, however, if the disc
subtends a larger solid angle as seen from the corona. Indeed,
doubling $\alpha$ to $0.2$ nearly doubles \dPhi in both energy
bands. A non-truncated accretion disc would be expected to be an
important contributor to the corona cooling, although, in this case,
$\epsilon \ll 1$ which would mitigate the impact of the reflection
contribution.



\subsection{Potential Implications of Time Dependent Effects}
\label{sub:timeDependence}
X-ray bursts increase in luminosity rapidly, reaching their peak flux
within a few seconds from onset \citep[e.g.,][]{KeekEtAl2018}, before, in most cases, decaying over the next several 10s of seconds. The \textsc{eqpair}
calculations from which we derive our results are time independent,
and predict an equilibrium spectrum and electron temperature at each
$L_{\mathrm{b}}/L_0$. This is valid because the Thomson time, a measure
of the timescale of photon-electron collisions, is \citep[e.g.,][]{padmanabhan2000theoretical,rybicki2008Book}
\begin{equation}\label{eq:tThomson}
t_\mathrm{Th} =\frac{1}{n_\mathrm{e}\sigma_\mathrm{T} c}= (2.2\times 10^{-5}\ \mathrm{s}) \left( {R \over
  10\ \mathrm{km}} \right ) \left ({\tau \over 1.5} \right )^{-1},
\end{equation}
much faster than the observed changes in X-ray luminosity. Therefore,
it is possible to make use of the results presented in
Sect.~\ref{sect:results} to consider how changes in the coronal
geometry during a burst would impact the emitted persistent spectrum.


For example, \citet{SimulatingtheCollapseofaThickAccretionDisk}
simulated a corona subject to an X-ray burst, and found that the
significant Compton cooling of the corona led to a loss of thermal
pressure support, causing the collapse of the corona into a thin
disc. Thus, in this scenario, where the burst did not cause
enhanced coronal dissipation, $\epsilon$ would decrease and $\tau$
increase during the burst. If the corona collapsed in this manner we
would expect the observed \frtwo (Eq.~\ref{eq:f3050}) to decrease, but
the extent of the drop would depend on the competing effects of
raising $\tau$ and lowering $\epsilon$. As we found that changes in
$\epsilon$ had a larger impact on the predicted spectrum (e.g.,
Fig.~\ref{varyTauEpsilonConstantB}) than $\tau$, it is likely
  that \frtwo would fall to a very low value if the corona entirely
  collapsed during an X-ray burst. The evolution of \frone
  (Eq.~\ref{eq:f824}) is more challenging to predict, as it is more
  impacted by the rising soft photon emission from the burst in addition to the drop
  in hard photon emission from the corona. The coronal electron temperature would
  initially decline due to flux of soft burst photons, but with the collapse
  of the corona fewer burst photons would intercept the corona,
  mitigating the cooling. Thus, depending on the peak temperature of
  the burst, \frone may show a moderate rise during the burst.

It may be possible that exceptionally powerful X-ray bursts, such as the photospheric radius expansion bursts \citep[e.g.,][]{Lewin1993}, or exceptionally long bursts, such as superbursts \citep[e.g.,][]{GallowayReview2017}, would disrupt the accretion flow to the point that the accretion rate is reduced during the burst and $\beta < 1$. In this case, the combined impact of a loss of coronal heating and intense Compton cooling would lead to severe hard X-ray shortages, likely through the $8$--$24$~\kev\ band, and the collapse of the corona would be a strong possibility. Further numerical work exploring these scenarios is needed to further understand the effects of these extreme bursts on the accretion disc and corona.

\section{Conclusions}
\label{sect:concl}


Analysis of the spectra of X-ray bursts provides an opportunity to
measure fundamental parameters of the heated neutron star, such as its
radius and compactness. However, the persistent spectrum produced by the accretion disc
and corona will also radiate during the burst and must be accounted
for in order to provide an accurate description of the
burst. Several analyses of X-ray bursts with hard X-ray instruments
have detected a hard X-ray shortage during the burst
\citep[e.g.,][]{Maccarone2003,jiEtAl2014KS1731-260,ji2014,Sanchez2020},
implying that the persistent spectrum is affected by the burst,
perhaps by Compton cooling due to the burst photons. This paper
presents the results of a series of \textsc{eqpair} calculations that
considered how burst-driven Compton cooling will effect the
temperature and emitted spectrum of a thermal accretion disc corona.
Our simulations show that the X-ray burst cools down the coronal
electrons and thereby will naturally lead to a hard X-ray shortage
that depends on the observed energy band and the geometrical
properties of the corona. Compton
cooling will almost always significantly reduce the flux in the
$30$--$50$~\kev\ band, but the emission in the $8$--$24$~keV band can
either increase or decrease depending on whether the high-energy tail of
the burst emission enters into the band. The magnitude of the flux
changes shows a strong dependence on the coronal aspect ratio, and a
weaker one on the optical depth across the corona. Therefore,
measurements of the hard X-ray flux deficit can constrain coronal
properties, independent of the exact geometry of the corona.

The size of the hard X-ray flux deficits predicted by the calculations
are broadly consistent with those measured or inferred from
observations of X-ray bursts
(Sect.~\ref{sub:comparisonObservations}). Therefore, detection of a
hard X-ray drop during a X-ray burst is consistent with a coronal
geometry that intercepts a significant fraction of burst photons, such
as a truncated inner accretion disc. As burst induced coronal cooling
always occurs, the lack of an observed hard X-ray deficit during a
burst would indicate a more compact coronal geometry that interacts
only weakly with burst photons.

Our calculations also considered the situation where the X-ray burst
increases the accretion rate during the burst, leading to enhanced
coronal dissipation (Sect.~\ref{sub:increasebeta}). In this case, the
increase in accretion energy into the corona during the burst somewhat
offsets the Compton cooling driven by the X-ray burst. As a result, the
hard X-ray flux deficits do not fall as low as when the accretion rate
was fixed. Careful and sensitive analysis of the hard X-ray emission
during X-ray bursts has the potential to provide critical information
on the physics of accretion during these violent episodes.

Compton cooling of the corona is generally expected during an X-ray
burst and should be accounted for in the analysis of X-ray burst
spectra. The cooling effects will change the shape of the spectrum (e.g.,
Fig.~\ref{spectraVaryingLbConstantB}), and therefore simply multiplying the
persistent spectrum by a factor during the burst does not accurately
capture the physical processes occurring within the corona. Both
changes in the normalization and spectral shape of the persistent
spectrum should be considered when fitting X-ray burst data in order
to reduce systematic errors on NS radius estimates \citep[e.g.,][]{Sanchez2020}.



Time dependent changes to the corona, although not strictly accounted for in
the calculations, may also be driven by the X-ray burst. Numerical
simulations of a burst interacting with a corona showed that
catastrophic Compton cooling could cause the collapse of the corona
\citep{SimulatingtheCollapseofaThickAccretionDisk}. Even if the corona
did not fully collapse, it is possible that the burst will drive
changes in $\epsilon$ and $\tau$ that would impact the predicted hard
X-ray flux. While it is possible to estimate the effects of such
changes using the equilibrium results presented here
(Sect.~\ref{sub:timeDependence}), a more thorough investigation of the
time dependent effects which makes use of the results of numerical
simulations is an important direction for future work.

\section*{Data Availability}
The data underlying this article will be shared on reasonable request
to the corresponding author.

\section*{Acknowledgements}
The authors acknowledge P.C. Fragile for helpful comments on a draft of the manuscript.




\bibliographystyle{mnras}
\bibliography{refs} 







\bsp	
\label{lastpage}
\end{document}